# An adaptive closed-loop ECoG decoder for long-term and stable bimanual control of an exoskeleton by a tetraplegic

Alexandre Moly[1*], Thomas Costecalde[1], Félix Martel[1], Christelle Larzabal[1], Serpil Karakas[1], Alexandre Verney[2], Guillaume Charvet[1], Stéphan Chabardes[3], Alim Louis Benabid[1], Tetiana Aksenova[1*]

[1] Univ. Grenoble Alpes, CEA, LETI, Clinatec, F-38000, Grenoble.
[2] Université Paris-Saclay, CEA, List, F-91120, Palaiseau, France.
[3] Centre Hospitalier Universitaire Grenoble Alpes, 38700, La Tronche, France.
[*] Correspondence and requests for materials should be addressed to F.M. (felix.martel@cea.fr) or T.A. (tetiana.aksenova@cea.fr).

**Abstract**
Brain-computer interfaces (BCIs) still face many challenges to step out of laboratories to be used in real-life applications. A key one persists in the high performance control of diverse effectors for complex tasks, using chronic and safe recorders. This control must be robust over time and of high decoding performance without continuous recalibration of the decoders. In the article, asynchronous control of an exoskeleton by a tetraplegic patient using a chronically implanted epidural electrocorticography (EpiCoG) implant is demonstrated. For this purpose, an adaptive online tensor-based decoder: the Recursive Exponentially Weighted Markov-Switching multi-Linear Model (REW-MSLM) was developed. We demonstrated over a period of 6 months the stability of the 8-dimensional alternative bimanual control of the exoskeleton and its virtual avatar using REW-MSLM without recalibration of the decoder.

**Introduction**
Brain-computer interfaces (BCIs) create a new communication pathway between the brain and an effector without neuromuscular activation. Among the various potential applications, the functional compensation/restoration of individuals suffering from severe motor disabilities has always been a focus for BCI research. Major milestones have been reached by the motor-BCI community over the years [1–3]. Nevertheless, many aspects need to be addressed to translate BCI-driven systems from laboratories into the patients' home for daily life applications.

The primary challenge of motor BCIs in clinical application is the high-dimensional control of effectors using safe, biocompatible and chronic neural recording systems. Brain signal recordings should remain stable and allow accurate neuronal signal decoding in conditions that are more demanding than in laboratories. The control of many degrees of freedom (DoF), up to 10,  based on microelectrode array (MEA) recordings  have been reported [1,3]. However, MEA recording systems are highly invasive, have biocompatibility issues and poor stability. They suffer from a decrease of signal-to-noise ratio over time [4,5], a high across-day variation in the neural signals[6,7] and still require wired recording systems despite recent efforts in this domain. Electrocorticography (ECoG) provides a good compromise between invasiveness and signal resolution[10–12]. Numerous pre-clinical and clinical studies demonstrated the interest in ECoG-based BCIs to control effectors[13–24] and highlighted the good signal-to-noise ratio and the stability of ECoG signals over months and even years[25–28]. Clinical results of high-dimensional (up to 8D) alternative bimanual control of a complex effector by a tetraplegic subject using epidural ECoG (EpiCoG) arrays have been recently reported [21]. This study outperformed the previously reported state-of-the-art ECoG-based BCIs with up to 3D control [13,14].

Another requirement to bring motor BCI into real life applications is to make the system act as a stand-alone device which can switch between idle/rest state (IS) and multiple active states (AS)[29,30] (referred to as asynchronous BCI). This is a critical point as the majority of the reported BCIs are synchronous cue-based action-oriented systems providing neuronal control of effector to the user during specific time intervals defined by an operator. For example, continuous BCI performance is often evaluated for single limb center-out experiments which classically reset the cursor position between trials without including the rest period. Besides not being representative of real life applications, it is likely that this may lead to unwanted activations/movements when the user does not intend to control the effector [29].

Moreover, single limb applications are limited compared to daily life actions which commonly require synchronized or alternative bimanual (or generally multi-limb) movements. Despite its clear benefit for patient motor deficit compensation and rehabilitation, multi-limb decoding has been poorly explored in the BCI field. Most breakthroughs involved the control of a single robotic arm or the movement decoding of one hand. Bimanual experiments have only been tested using MEAs with virtual effectors [31] or during ECoG-based movement detection experiments [20] in non-human primates. While bimanual and/or asynchronous BCIs are not very common, several decoding strategies have been proposed. In particular, the problem of multi-finger movement trajectory reconstruction from ECoG recordings was studied. In most of the cases, hybrid models were employed by mixing classifier outputs to detect finger activations and continuous decoders to predict their respective movements [32–34].

A BCI decoder must be sufficiently optimized to enable computation time suitable for real-time application. Despite promising results, translating the off-line trajectory reconstruction algorithm to real-time closed-loop experiments is generally a challenging task. Even if a decoder meets the real-time requirements, drops in the decoding performance have been reported repeatedly when decoders calibrated off-line using open-loop experiments were used for online decoding[4,35,36]. Open-and closed-loop model training lead to distinct decoders[37]. To take the patient feedback into account, BCI studies employ adaptive decoders which integrate the decoding model parameters identification into closed-loop BCI sessions. Adaptive decoders update their parameters in an incremental manner with new incoming data, optimizing the model parameters in real time. While several adaptive linear and nonlinear regression and classification decoders were proposed for MEA [7,38–42] and electroencephalography (EEG) [43–47] driven BCI, only a few adaptive decoders were developed for ECoG recordings [19]. Most adaptive algorithms are restricted to linear decoders, which may be limiting for complex effector control with a high number of DoF. Closed-loop adaptive model calibration is one strategy to create model stable over time[7,36] which is a major challenge, considering the non-stationarity and the intra-subject variability of the brain's signals (inattention, habituation). So far, stable long-term BCIs were only achieved using non-adaptive brain switch decoders (1D) for a period of 4 months with local field potentials[48] and 36 months with ECoG recordings [27].

In the present article, the Recursive Exponentially Weighted Markov-Switching multi-Linear Model (REW-MSLM) is proposed to address the lack of stable asynchronous BCI for bimanual/whole-body effector control from EpiCoG recordings in tetraplegics[53]. The tensor-based piecewise linear REW-MSLM algorithm is an online adaptive supervised learning algorithm. It updates a decoder in real time in an incremental manner during the calibration period of closed-loop BCI experimental sessions. In this article we report the case study of a closed-loop 8D control performed by a tetraplegic patient in an exoskeleton and with a virtual avatar. These results outperform the 3D control of previous ECoG-based state-of-the-art BCIs [13,14,18]. We demonstrated the remarkable stability of the BCI system and the stable performance of the REW-MSLM decoder which was not recalibrated for more than 5 months when the patient was in the exoskeleton and for more than 6.5 months using the virtual avatar. For both effectors, the patient was able to switch reliably between discrete states and demonstrated relevant control for continuous movements. The decoding performance outperformed the ECoG-based BCIs state-of-the-art for which such a long-term robustness was never reported before. Compared to the classic center-out experiments the patient was able to perform more complicated tasks such as multiple alternative point-to-point pursuit tasks

**Methods**

REW-MSLM decoder.
The Recursive Exponentially Weighted Markov-Switching multi-Linear Model (REW-MSLM) is an online tensor-based fully adaptive mixture of multi-linear experts algorithm (Figure 1). The REW-MSLM inherits the Markov-switching linear model (MSLM)[17] mixture of experts (ME) structure, generalizing the MSLM model to tensor-input tensor-output variables and introducing the recursive model parameter identification procedure inspired by the Recursive Exponentially Weighted N-way Partial Least Squares (REW-NPLS) method[19].

*MSLM description.*
The MSLM[17] is a hybrid discrete/continuous decoder based on a ME model structure. A ME mixes or switches independent decoders, called "experts". Basic assumption of ME is that each expert decodes its own specific region of feature space[54]. Experts are mixed according to the "gating" model which estimates the probability of an expert to be activated or inhibited. This probability is used to compute gating coefficients to weight experts' outputs. Additionally, MSLM uses dynamic gating assuming a hidden Markov model (HMM) for the state sequence to improve the decoder robustness. Conventional MSLM is vector input vector output model and employs linear experts. Both experts and gating models are identified offline. Application of MSLM in motor BCI was limited to offline studies: 3D-trajectory decoding of single limb wrist from nonhuman primates (NHP) ECoG with 2 states separating rest and movement periods and 1D-trajectory decoding of fingers movement with states associated to individual finger activation and rest periods from ECoG recordings of abled body patients undergoing pre-surgical evaluation[17].

*REW-NPLS description*
Due to the robustness in the computation of high dimensional data, algorithms of the Partial Least Squares (PLS) family were frequently used in continuous and discrete BCI decoding experiments. Numerous publications which reported offline ECoG-based hand trajectory decoding[18,20,22,55–57], and EEG-based classification or cursor decoding[58,59] confirmed the interest for such algorithms. The classical PLS regression algorithm is an offline procedure based on the iterative projection of input $\mathbf{x}_t \in \mathbb{R}^m$ and output $\mathbf{y}_t \in \mathbb{R}^n$ variables into a latent variable space of dimension $f$ ($f$ is referred as the PLS hyperparameter). Projectors are estimated by maximizing the covariance between the input and the output latent variables[60]. The subspace dimension $f$ is typically determined through cross-validation. For online modeling, recursive PLS (RPLS) and recursive exponentially weighted PLS (REW PLS)[61–63] were proposed.

A generalization of the conventional PLS algorithm to tensor data: the N-way Partial Least Square (NPLS) algorithm, was proposed by Bro[64,65]. A tensor is a generalization of a matrix to higher order dimensions, also known as ways or modes. Vectors and matrices are special cases of tensors with one and two modes respectively[66]. Tensor-based algorithms emerged as a promising strategy in the BCI field. They allowed simultaneous processing of high-dimensional data in the temporal, frequency and spatial domains[19,66]. The NPLS algorithm projects the input and output tensors into low dimensional space of latent variables using a low rank tensor decomposition. It improves the stability and robustness of the model compared to the classic unfold PLS leading to more accurate and interpretable predictions[64,65] while preserving the structure of the data.

For the online tensor data flow modelling, the Recursive N-way PLS (RNPLS)[61] which is a generalization of the RPLS algorithm to tensor variables, and the recursive exponentially weighted N-way PLS (REW-NPLS)[19] inspired from kernel recursive exponentially weighted PLS (REW PLS)[61–63] were proposed. RNPLS still requires fixing the hyperparameter $f$ from offline preliminary study whereas the Recursive-Validation procedure used in REW-NPLS for online optimization of the hyperparameter $f$ enables a fully adaptive algorithm [19]. The decoder is entirely tuned in real time. The Kernel REW-NPLS is also more computationally efficient than the RPLS algorithm.

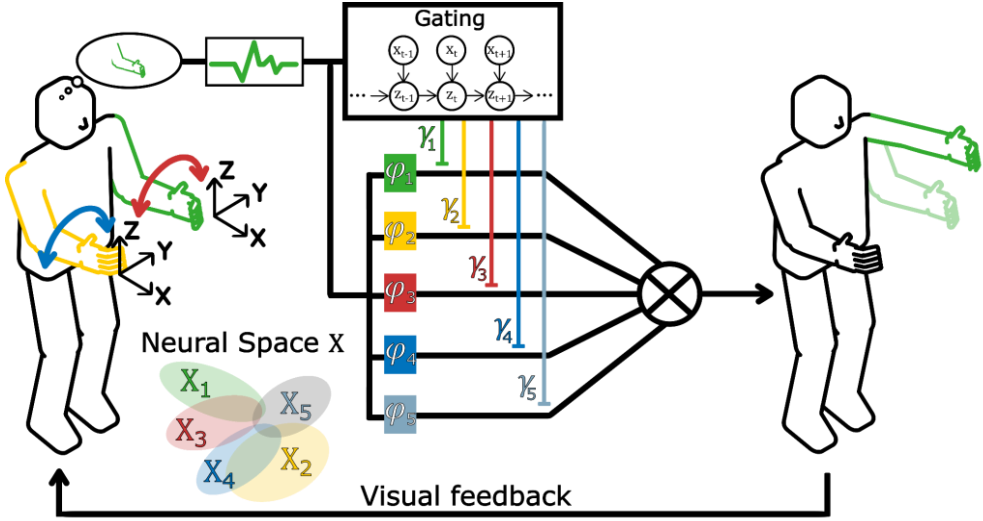

*Figure 1:* ***Recursive exponentially weighted Markov-switching linear model (REW-MSLM) architecture.*** *The REW-MSLM includes a mixture of experts model, which can be described as the parallel computation of several predictions from different regression models (experts) that are weighted (enhanced or inhibited) according to the input variables using a classifier (gate). We hypothesize that the input feature space $X$ can be divided into several specific local regions $X_k$ and that each sub-space can be fitted using local multilinear functions $\varphi_k$ associated with an expert. Multilinear functions $\varphi_k$ are estimated using $k$ independent REW-NPLS models. The selected expert is determined based on the dynamic gating model. The gating model is a hidden Markov model (HMM) which computes the probability $\gamma_k$ for each expert to be activated. Commands are decoded by the REW-MSLM and sent to the effector to provide visual feedback to the patient.*

*REW-MSLM description*

The REW-MSLM inherits the MSLM mixture of experts structure. A basic assumption of ME is that each expert decode its own specific region of feature space. Given $\underline{\mathbf{X}}_t \in \underline{X} \subset \mathbb{R}^{I_1 \times \ldots \times I_m}$ and $\underline{\mathbf{Y}}_t \in \underline{Y} \subset \mathbb{R}^{J_1 \times \ldots \times J_n}$ the independent and dependent $m$ and $n$ order tensor variables at time $t$ respectively, the feature space of independent variables is supposed to be partitioned into $K$ possibly overlapping regions $\underline{X} = \cup_{k=1}^{K} \underline{X}_k$. It is assumed that the space of input variables is mapped to the space of output variables using $K$ local multilinear functions $\Phi = \{\varphi_k: \underline{X}_k \to \underline{Y}, \ k = 1, 2, \ldots, K\}$. Let $z_t \in [1; K] \subset \mathbb{N}^*$ be a latent state variable which defines the selected local multilinear function at time $t$ (expert) : $\underline{\mathbf{Y}}_t = \varphi_{z_t}(\underline{\mathbf{X}}_t)$. Similar to MSLM, dynamic gating is introduced using a first-order HMM[17]. The latent state variable $z_t$ is assumed to follow the first order Markovian assumption, which states that the dependence of $z_t$ is limited to the past state $z_{t-1}$. $\underline{\mathbf{Y}}_t$ is estimated as follows:

$$\widehat{\underline{\mathbf{Y}}}_t = \sum_{k=1}^{K} \gamma_{k,t} \left( \underline{\mathbf{Beta}}_k \, \underline{\mathbf{X}}_t + \underline{\mathbf{bias}}_k \right).$$

Here, $\underline{\mathbf{Beta}}_k$ and $\underline{\mathbf{bias}}_k$ are the $k^{th}$ expert's tensor parameters and its associated bias. $\gamma_{k,t} = p(z_t = k | \underline{\mathbf{X}}_{1:t})$ is the dynamic gating weight coefficient associated with the $k^{th}$ expert at time $t$. REW-MSLM models are entirely defined through the experts' parameters $\theta_e = \{\underline{\mathbf{Beta}}_k, \underline{\mathbf{bias}}_k\}_{k=1}^{K}$ and HMM parameters $\theta_g = \{\mathbf{A}, \{d_k\}_{k=1}^{K}, \boldsymbol{\pi}\}$, where $\mathbf{A}$ is the transition matrix, $\mathbf{A} = (a_{ij}) \in \mathbb{R}^{K \times K}$, $a_{ij} = p(z_t = j | z_{t-1} = i)$, $\{d_k\}_{k=1}^{K}$ is the set of parameters employed to estimate conditional emission probability of the observed variables $p(\underline{\mathbf{X}}_t | z_t)$, and $\boldsymbol{\pi} \in \mathbb{R}^K$ is the initial state probability vector at $t = 0$.

REW-MSLM online/incremental training.

The proposed REW-MSLM algorithm recursively estimates $\Theta = \{\theta_g, \theta_e\}$ with a supervised training procedure. At each update $u$, the corresponding block of training dataset $\{\underline{\mathbf{X}}_u, \underline{\mathbf{Y}}_u, \mathbf{z}_u\}$ is given with $\underline{\mathbf{X}}_u \in \mathbb{R}^{\Delta L \times I_1 \times \ldots \times I_m}$, $\underline{\mathbf{Y}}_u \in \mathbb{R}^{\Delta L \times J_1 \times \ldots \times J_n}$, $\mathbf{z}_u = (z_{t_1}, \ldots, z_{t_1 + \Delta L})^T \subset \mathbb{N}^{*\Delta L}$ and $\Delta L$ the update block size. The $K$ local multilinear functions $\varphi_k$ are estimated using expert's specific samples. The $k^{th}$ expert's parameter update is performed on the training dataset $\{\underline{\mathbf{X}}_u^k, \underline{\mathbf{Y}}_u^k\}$. $\underline{\mathbf{X}}_u^k$ and $\underline{\mathbf{Y}}_u^k$ are sub-tensors of $\underline{\mathbf{X}}_u$ and $\underline{\mathbf{Y}}_u$ formed by samples labelled as belonging to state $k$. The $k^{th}$ expert's parameters are updated using the REW-NPLS algorithm: REW-NPLS$_e$ = REW-NPLS$(\underline{\mathbf{X}}_u^k, \underline{\mathbf{Y}}_u^k)$ with the forgetting factor $\lambda_k$, $0 \leq \lambda_k \leq 1$.

For online optimization latent variable space dimension (hyperparameter $f$), the REW-NPLS$_e$ algorithm estimates a set of $F$ models for each expert $\{\underline{\mathbf{Beta}}_{u,k}^f, \underline{\mathbf{bias}}_{u,k}^f\}_{k,f=1}^{K,F}$. $F \in \mathbb{N}^*$ is the fixed highest latent space dimension. The optimal hyperparameter of the $k^{th}$ expert $f_k^* \leq F$ is selected following the Recursive-Validation procedure[19]. For the currently available models, the Recursive-Validation exploits the newly available $\{\underline{\mathbf{X}}_u^k, \underline{\mathbf{Y}}_u^k\}$ dataset as testing data to evaluate the best hyperperparameters before this dataset is used as a training dataset for the models updating. The best models are chosen independently for each expert: $\{\underline{\mathbf{Beta}}_k, \underline{\mathbf{bias}}_k\}_{k=1}^K = \{\underline{\mathbf{Beta}}_{u,k}^{f_k^*}, \underline{\mathbf{bias}}_{u,k}^{f_k^*}\}_{k=1}^K$, and are used for real-time decoding of the neural signals.

Similarly, at each update $u$, the HMM gating parameter are updated using the update block dataset $\{\underline{\mathbf{X}}_u, \mathbf{z}_u\}$. The HMM transition matrix $\mathbf{A}$ is approximated by counting the successive transition of states in $\mathbf{z}_u$ and is weighted with the transition matrix estimated during the previous updates with the forgetting factor $\lambda_g$, $0 \leq \lambda_g \leq 1$. The HMM conditional emission probabilities $p(\underline{\mathbf{X}}_t|z_t)$ are inferred through the combination of $p(z_t|\underline{\mathbf{X}}_t)$ and class prior $p(z_t)$ using the Bayes' theorem[67]. The REW-NPLS discriminative decoder is embedded into the HMM-based gating process to evaluate $p(z_t|\underline{\mathbf{X}}_{1:t})$. REW-NPLS was used because of its online adaptive characteristics and its relevance for high dimensional input variable. A discriminative decoder is selected instead of generative ones due to benefits for high dimensional and complex dependencies of features [68,69]. The decoder is trained on the observation tensor of input variables $\underline{\mathbf{X}}_u$ and the latent state dummy variable matrix $\mathbf{Z}_u \in \{0,1\}^{K \times \Delta L}$ where the column-wise (single) non-zero element depicts the activated state for each sample (one-hot encoding).

The discriminative REW-NPLS decoder computes a set of $F$ multilinear models $\{\underline{\mathbf{B}}_u^f, \mathbf{b}_u^f\}_{f=1}^F$, where $\underline{\mathbf{B}}_u^f \in \mathbb{R}^{K \times I_1 \times \ldots \times I_m}$ and $\mathbf{b}_u^f \in \mathbb{R}^K$ are the tensor of the gating model parameters and its related bias. The Recursive-Validation procedure selects the best model based on the estimated gating optimal hyperparameter $f_g^* \leq F$ and defines the optimal gating model as $\{\underline{\mathbf{B}}, \mathbf{b}\} = \{\underline{\mathbf{B}}_u^{f_g^*}, \mathbf{b}_u^{f_g^*}\}$ for the dynamic gating weight $\gamma_{k,t}$ estimation. The output variable $\hat{\mathbf{z}}_t \in \mathbb{R}^K$ determines how likely each hidden state is generated based on $\underline{\mathbf{X}}_t$. The prediction $\hat{\mathbf{z}}_t$ is computed from the discriminative REW-NPLS decoder. Then, $p(z_t|\underline{\mathbf{X}}_t)$ is evaluated with the softmax function[54] to compute $\gamma_{k,t} = p(z_t|\underline{\mathbf{X}}_{1:t})$ using HMM forward algorithm.

REW-MSLM uses dynamic HMM gating. The equivalent mixture of expert algorithm using static gating (without HMM) is referred as REW-SLM. REW-SLM gating is computed with the REW-NPLS trained on explanatory variables and latent states, using the softmax function but without the HMM forward algorithm.

REW-MSLM application.

In real time, each expert $\{\underline{\mathbf{Beta}}_k, \underline{\mathbf{bias}}_k\}_{k=1}^K$ output is estimated for each new input data buffer after feature extraction $\underline{\mathbf{X}}_t$. The dynamic gating coefficients $\gamma_{k,t}$ are estimated using the latent state variable estimator $\hat{\mathbf{z}}_t$ post-processed with a softmax function[54] and the HMM forward algorithm[70]. The forward algorithm evaluates $\gamma_{k,t}$ by considering the past and current observations:

$$\hat{\mathbf{z}}_t = \underline{\mathbf{B}}\,\underline{\mathbf{X}}_t + \mathbf{b},$$

$$p(z_t = k|\underline{\mathbf{X}}_t) = \frac{\exp(\hat{z}_{k,t})}{\sum_{i=1}^K \exp(\hat{z}_{i,t})},$$

$$p(z_t = k, \underline{\mathbf{X}}_{1:t}) = p(\underline{\mathbf{X}}_t|z_t = k) \sum_{j=1}^K a_{kj}\, \gamma_{k,t-1},$$

$$\gamma_{k,t} = p(z_t = k|\underline{\mathbf{X}}_{1:t}) = \frac{p(z_t = k, \underline{\mathbf{X}}_{1:t})}{\sum_{j=1}^K p(z_t = k, \underline{\mathbf{X}}_{1:t})}.$$

Clinical Trial description

The REW-MSLM algorithm was tested and applied as the neural signal decoder during the "BCI and Tetraplegia" clinical trial (ClinicalTrials.gov, NCT02550522[71,72]). The clinical trial was approved by the French authorities: National Agency for the Safety of Medicines and Health Products (Agence nationale de sécurité du médicament et des produits de santé: ANSM), registration Number 2015-A00650-49, and the ethical Committee for the Protection of Individuals (Comité de Protection des Personnes - CPP), registration number 15-CHUG-19. All research activities were carried out in accordance with the guidelines and regulations of the ANSM and the CPP.

The REW-MSLM was tested on a single patient. The patient signed informed consent prior to surgery. Details of the clinical trial protocol are available in [21]. The subject was a 29-year-old right-handed male with traumatic sensorimotor tetraplegia caused by a complete C4–C5 spinal cord injury 2 years prior to the study. The patient can perform neck, shoulder and small upper limb movements by contraction of the biceps at the elbow and extensors of the wrists. American Spinal Injury Association Impairment (ASIA) scores the contraction of the biceps at the elbow at 4 and 5 for the right and left body side respectively, whereas extensors contractions were scored at 0 and 3 for the right and left wrists, respectively. With the exception of the cited muscles, all others muscles below were scored 0 on the ASIA scale. Moreover, the sensory-motor deficit was complete.

The patient underwent bilateral implantation of two chronic wireless WIMAGINE implants [21,53] for EpiCoG signal recording on June 21, 2017 and since, underwent training for more than 28 months. Two WIMAGINE recording systems were implanted into the skull within a 25 mm radius craniotomy placed in front of the sensory motor cortex (SMC). The electrodes located at the implant lower surface are in contact with the dura mater. Before surgery, the subject's SMC was localized clearly using functional imaging (fMRI and MEG) as the subject imagined virtual movements of his limbs or performed real motor tasks when possible. Details are provided in[21]. WIMAGINE is an active implantable medical device composed of 64 plane platinum iridium 90/10 electrodes with a 2.3 mm diameter and 4-4.5 mm inter-electrodes distance on the lateral and antero-posterior directions[26] dedicated to ECoG neural signal recordings. WIMAGINE implant was shown to be safe for long-term EpiCoG signal recording [26,53]. The EpiECoG signals are low and high pass filtered in a bandwidth from 0.5Hz to 300Hz using analog low pass filters as well as a digital low pass FIR filter directly embedded into the implant hardware[53]. The digitized EpiCoG data are radiotransmitted to a custom designed base station connected to a computer[53]

Since the implantation date, the patient was trained using a custom-made BCI platform to control multiple real and virtual effectors[21] (Figure 2). The article presents a series of experiments performed in the laboratory with the EMY (Enhancing MobilitY)[21,73] exoskeleton and with EMY's virtual avatar replica for training at home. EMY is a wearable fully motorized four-limb robotic neuroprosthesis (14 joints, 14 actuated DoFs) equipped by a computer station receiving radio-emitted EpiCoG signals. After decoding, the neuronal signal is translated into the motor commands which activates the limbs and joints to produce adequate movements, mimicking natural limbs movements.

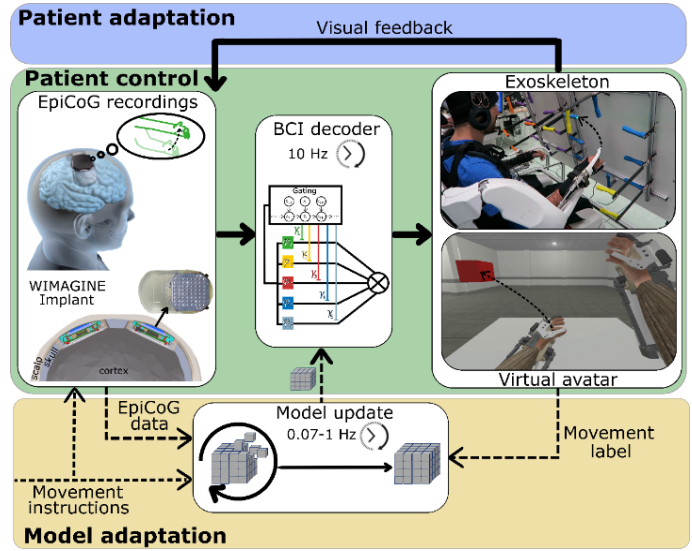

*Figure 2: **BCI platform for the "BCI and tetraplegia" clinical research protocol in CLINATEC**. Two wireless WIMAGINE implants with a 64-electrode array [53] are used to record EpiCoG signals which are radio-transmitted to an external processing unit. Implants were placed into the skull in contact with the dura mater above the motor cortex by a craniotomy. EpiCoG recordings are sent to the BCI decoders which translate the neural signals into decisions (at 10Hz frequency) to control various effectors. The exoskeleton is used for training in CLINATEC and the virtual avatar is used to train the patient at home. Both effectors provide a visual feedback to the patient to adapt and respond in a closed-loop fashion to model predictions. The EpiCoG data and the movement instructions are used to update the model in real time during the closed-loop BCI calibration sessions. The model is updated at a 0.07-1Hz frequency.*

Experimental setup

In this study, experiments performed between March 5th, 2018 and May 19th, 2019 are considered. Experiments were carried out in the laboratory three successive days per month. During these sessions, the patient was strapped into the EMY exoskeleton. For the remaining weeks, experiments were performed in the patient's home three days a week. The patient was installed in his wheelchair in front of the computer screen (Figure 2). During all the experiments, the patient was allowed to move and talk freely during the training and test sessions in order to create models that are robust to artefacts related to muscular activities such as head movements. During the experimental sessions, 32 electrodes for each implant were selected in a checkerboard-like pattern because of temporarily limited data rates, caused by restricted radio link.

All the experiments were online closed-loop BCI experiments. Effectors were controlled using the patient's neural activity at a 10 Hz frequency rate.

The experiments were divided into two phases. The training phase (optional) was designed for online updates of the REW-MSLM decoder. The test phase, during which the model was fixed, was used for the performance evaluation. At the beginning of each session, the decoding model was initialized to zero or was set to model parameters determined from previous sessions. A support/assistance system was optionally provided to the patient during the early model calibration phase, if the decoding model was created from scratch. The assisted control command $\mathbf{y}_t^{assist}$ provided during early model calibration phase is defined as: $\mathbf{y}_t^{assist} = \omega_c\, \hat{\mathbf{y}}_t + \omega_s\, \mathbf{y}_t$. Here, $\hat{\mathbf{y}}_t$ is the decoder prediction, $\mathbf{y}_t$ is the optimal prediction, $\omega_c$ is the patient's control weight and $\omega_s$ is the support provided to the model at the beginning of the training phase, $\omega_s = 1 - \omega_c$. A maximum of 30% assistance was provided. All performance evaluations were computed on unassisted test experiments out of calibration (update) periods (Figure 3).

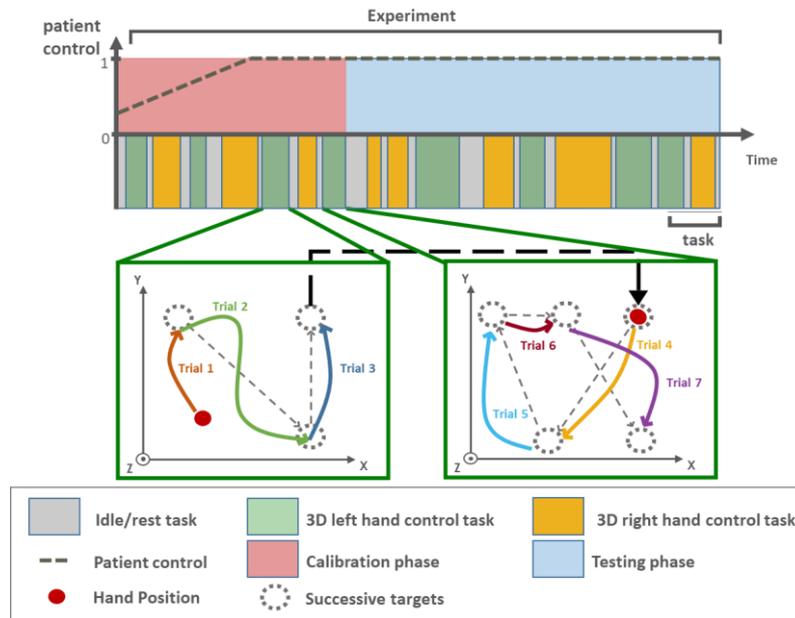

*Figure 3: Examples of 6D alternative multi-limb pursuit tasks. One session is composed of successive tasks. Each active task is composed of several trials in which the 3D cursor must reach the proposed targets. The cursor position is not reset between tasks, during task and during idle state.*

Two types of asynchronous alternative high-dimensional controls of the two arms are presented in the article: 1/ two-handed reaching tasks in 3D with the virtual avatar (6D control) ; 2/ two-handed reaching in 3D plus 1D wrists rotation performed in the exoskeleton or with the virtual avatar (8D control).

The experimental 6D-control sessions were previously performed during the clinical trial. During these experiments, the generic REW NPLS algorithm was employed for real time BCI control. In the current article the 6D dataset was used for offline comparison between the generic REW-NPLS algorithm and REW-MSLM.

Sessions with the patient performing 8D control were carried out using proposed REW-MSLM algorithm.

Control tasks
In both experimental series, each session was composed by successive tasks decided by an experimenter (Figure 3). Each task corresponded to one of the available states, the idle state (IS) or the active states (AS). During the IS, no target was presented to the patient and the patient had to remain in a non-active state until the next instruction. The AS tasks corresponded to the translation of the left ($AS_{LH}$) and right ($AS_{RH}$) hand in the 3D space and to the 1D angular rotation of the left ($AS_{LW}$) and right ($AS_{RW}$) wrist.

The experimenter asked the patient to perform the mental tasks using upper limb motor imagery. The patient was allowed to define his strategy consisting of a combination of arms, wrists and fingers movements. After choosing a strategy, he was urged to maintain it constant for each task through the experiments.

Patient's performance was evaluated using point-to-point pursuit tasks (Figure 3). Each task was made of several successive trials during which the patient attempted to reach a target location which was set sequentially with his left/right hand or to rotate his left/right wrist up to a specific angle value. During a session, the hand position was not reset by the system between the different states, tasks and trials. For a given AS, the starting position of the hand for a trial was the position of the hand at the end of the previous trial of the same AS. An illustration, a session with the three IS, $AS_{LH}$ and $AS_{RH}$ is shown in the Figure 3. A total of 22 targets symmetrically distributed in two 3D cubes (11 targets per hand) were proposed to the patient.

REW-MSLM decoder integration.
In order to perform online decoding with online closed-loop decoder adaptation, the main application loop for the online decoding and the adaptation loop for the update of the REW-MSLM submodels were split and implemented in two independent processes/threads while communicating through shared memory.

The application loop received the data from the WIMAGINE implants and decoded the neural signals in order to control the exoskeleton. In order to incrementally update the REW-MSLM decoder, the input and output features were stacked in buffers before to be sent to the calibration loop in order to perform the incremental batch update of the gate and expert models.

In this study, neural signals were recorded at a 586 Hz sampling rate and were decoded at 10 Hz while the model was updated at a 0.07Hz update rate (every 15 seconds). Therefore, each incremental update was based on $\Delta L = 150$ samples. Every analysis and online experiment, including training and decoding, were performed with Matlab2017b using an Intel Xeon E5-2620v3 computer with 64 GB RAM.

*REW-MSLM parameters and structure*
REW-MSLM states were associated to particular tasks. In the present study, a ME structure with 3 states: idle (IS ), left ($AS_{LH}$) and right ($AS_{RH}$) hand translation states, was considered in an offline comparison study to decode asynchronous alternative 3D two-hand reaching tasks. A ME structure with 5 states: idle (IS), left ($AS_{LH}$) and right ($AS_{RH}$) hand translation, left ($AS_{LW}$,) and right ($AS_{RW}$) wrists rotation states was used during the online closed-loop experimental sessions using the REW-MSLM algorithm to control the exoskeleton or the virtual avatar.

*Neuronal feature extraction*
During the experimental sessions, at each time step $t$, EpiCoG epochs of neural signals for all the electrodes, $\mathbf{X}_t \in \mathbb{R}^{586 \times 64}$, were generated using a $\Delta t = 1\ s$ window with 100 ms sliding step[19]. ECoG epochs were mapped to the temporal frequency space using a complex continuous wavelet transform (CCWT) (Morlet) with a frequency range from 10 to 150 Hz (10 Hz step) for all the electrodes. CCWT is a feature extraction strategy that was widely used in the field of BCIs. Its efficiency has previously been demonstrated[16,17,19,20,22]. The absolute value of CCWT was decimated along the temporal modality to obtain a 10-point description of a 1s time epoch for each frequency band and for each channel, resulting in the temporal-frequency-spatial neural feature tensor $\underline{\mathbf{X}}_t \in \mathbb{R}^{10 \times 15 \times 64}$.

*Output feature extraction*
REW-MSLM is a supervised learning algorithm. Movement (output) features are extracted for model training during calibration/update period. At the time $t$ the optimal continuous movement $\mathbf{y}_t$ and the discrete state labels $z_t \in [1; K] \subset \mathbb{N}^*$, where $K$ is the number of states, were estimated. $\mathbf{y}_t = ((\mathbf{y}_t^{Ltr})^T, (\mathbf{y}_t^{Rtr})^T)^T$, $\mathbf{y}_t \in \mathbb{R}^6$, for alternative two-handed 3D reaching tasks and $\mathbf{y}_t = ((\mathbf{y}_t^{Ltr})^T, y_t^{Lr}, (\mathbf{y}_t^{Rtr})^T, y_t^{Rr})^T$, $\mathbf{y}_t \in \mathbb{R}^8$, if 1D wrists rotation is additionally considered. Here $\mathbf{y}_t^{Ltr} \in \mathbb{R}^3$ and $\mathbf{y}_t^{Rtr} \in \mathbb{R}^3$ are left and right hand translation components of $\mathbf{y}_t$. They are defined as the 3D Cartesian vector between the current hand position at the time moment $t$ and the target position. $y_t^{Lr} \in \mathbb{R}$ and $y_t^{Rr} \in \mathbb{R}$ are left and right wrist rotation components of $\mathbf{y}_t$, defined as a 1D angle between the current angle position and the target angle position (Figure 4)[42]. The discrete state $z_t$ labels are determined by the task instruction. $K = 3$ in the 6D control experiments (idle state, left hand translation, and right hand translation states) and $K = 5$ (idle state, left hand translation, right hand translation, left wrist and right wrist rotation states) in the 8D experiments. Output features were extracted during experiments at 10 Hz (Figure 4).

The decoder prediction ($\hat{\mathbf{y}}_t \in \mathbb{R}^6$ for 6D experiments, and $\hat{\mathbf{y}}_t \in \mathbb{R}^8$ for 8D experiments) sent to the exoskeleton after post-processing is defined as the Cartesian position increments for 3D hand translation and as angular increments for 1D wrist rotation. $\hat{\mathbf{y}}_t$ is post-processed by the exoskeleton control system using inverse kinematics to transform the Cartesian prediction into joint movements.

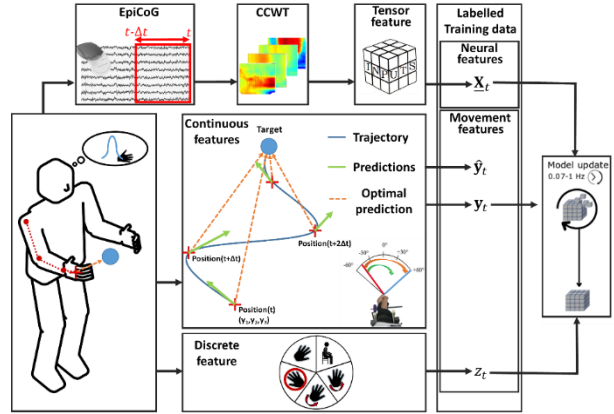

*Figure 4: Feature extraction for supervised training procedure. Neural and movement features recorded during the closed-loop experiments were used for the adaptive supervised training procedure based on the temporal-frequency-spatial neural feature tensor $\underline{X}_t$. The optimal predicted direction $y_t$ is defined as the 3D Cartesian vector between the current position and the target position for the 3D hand translation and as the 1D angular vector between the current angle and the target angle for 1D wrist rotation. The discrete state labels is noted $z_t$. At each time step $t$, the neural activity is acquired. The last second is mapped to the spatial frequency space using a CCWT to create tensor-shaped neural features. Simultaneously, the prediction from the current model $\hat{y}_t$, the optimal prediction $y_t$ according to the current position and the associated state $z_t$ are recorded as movement features. $\underline{X}_t$, $y_t$ and $z_t$ are stored in a buffer until the next update (every 15 s) of the REW-MSLM decoder.*

REW-MSLM performance evaluation

Two series of online asynchronous BCI experiments described above were performed to highlight REW-MSLM performance. Experimental sessions of alternative 3D two-hand reaching tasks (6D) of the virtual avatar effector were used for the offline/pseudo-online models comparison. The datasets were recorded during online closed-loop experiments using REW-NPLS decoder previously integrated to the BCI system. They were re-computed with different algorithms in a pseudo online manner using sample-wise indicators for the performance evaluation. Pseudo-online simulation was conducted using the same parameters (buffer size, batch training etc.) and the same model application procedure as the one used for online real-time experiments to reproduce the online experiment conditions. Pseudo-online comparison is not fully generalizable for the online case. Nevertheless, it allows characterizing to some extent the algorithms before an integration into the clinical BCI decoding platform. Finally, the REW-MSLM algorithm was integrated into the BCI platform to carry out 8D control experiments with the patient.

*Offline comparison*

The REW-MSLM is a ME algorithm which mixes discrete (state discrimination) and continuous (experts) decoding. We first highlighted the discrete multi-state decoding performance for an asynchronous control paradigm evaluating the accuracy of switching between all active states (AS) and, especially, the robustness of idle state (IS) support. REW-MSLM discrete decoding was compared to REW-NPLS algorithm thresholded in post-processing (referred to as REW-NPLS$_T$) to label the continuous decoding results as discrete IS and AS states. Such a comparison underlined the benefits of computing an additional discrete decoder to inhibit the experts continuous outputs. Next, the REW-MSLM was compared to its own variant without HMM (REW-SLM) to determine the benefits of dynamic HMM gating.

The REW-MSLM algorithm benefits from the ME structure which splits the neural space into state-related subsets associated to independent expert decoders. The training data are divided into subsets associated with particular experts, allowing independent expert learning. However, continuous decoder-experts are trained on a smaller specific subset of the training dataset. This may affect regression performance. The expert-specific subset training strategy was evaluated by comparing the continuous decoding performance of piece-wise linear REW-MSLM to state-of-the-art adaptive linear regression which was trained on the entire dataset. REW-MSLM experts trained on specific subsets of the training dataset were compared to the REW-NPLS model trained on the entire dataset.

Performances were compared to the REW-NPLS algorithm because it is a state-of-the-art online adaptive tensor-input tensor-output algorithm which has been previously used for closed-loop ECoG-based BCI [19,21]. A single multilinear decoder was identified by REW-NPLS. A ME structure with 3 states: idle (IS), left ($AS_{LH}$) and right ($AS_{RH}$) hand translation states, was considered using REW-MSLM.

Both discrete (gating) and continuous (experts) decoders were evaluated on three different experimental paradigms. For paradigm A, the decoder was calibrated from scratch at the beginning of each session with a small training dataset. Sessions during the paradigm A ($n = 5$) were self-contained experiments. The models were independently created (initialized to zero), trained and tested during the same experiment. Model adaptation with multiple calibration sessions was studied with the paradigm B. Sessions in series B ($n = 4$) were performed to evaluate the importance of cross-session training. The models were initialized to zero in the first session. Then, the models created during the previous sessions were used to initialize the model parameters of the next session. Finally, the last model created during experimental series B were used without adaptation (paradigm C). The C series of experiments ($n = 5$) were performed to evaluate model robustness across time and were carried out from 9 days to 28 days after model calibration. All experiments are closed-loop sessions recorded between March and June 2018.

*Offline Performance criteria*
Discrete performance was evaluated using the accuracy ($acc$) and the F-score ($fsc$) (see Appendix) for the multi-class case (IS versus $AS_{LH}$ versus $AS_{RH}$) and the two binary cases: IS versus $AS_{LH}$ and $AS_{RH}$ combined (named AS) and the classification between active states ($AS_{LH}$ versus $AS_{RH}$). Accuracy and F-score indicators are sample-based performance estimators and do not reflect the dynamic behaviour of misclassified samples. Consecutive misclassified samples were counted to evaluate the error block rate ($ErrB_{rate}$) and the error block durations ($ErrB_{time}$). A sequence of misclassified samples is referred as an error block. Error block durations present mean duration of error blocks in seconds. The error block rate represents the occurrence of blocks of wrong detections per minute. Additionally, the latency ($lat$) between the task instruction initiated by the experimenter and initiation of the movement by the patient was computed to evaluate the response time variation introduced by the HMM. The computed latency includes the patient's reaction time and the decoder latencies.

The trajectories performed during the online closed-loop experiments are related to the decoding model currently used during the experiments and patient's feedback. Therefore, they cannot be use to evaluate the performance of different algorithms in pseudo-online simulation.

Continuous performance comparison was evaluated using sample-wise Cosine Similarity ($CosSim$) index, the averaged normalized dot product of the predicted $\hat{\mathbf{y}}_t$ and the optimal $\mathbf{y}_t$ (see Appendix). Continuous performance was evaluated for the left hand ($CosSim_{LH}$) and the right hand ($CosSim_{RH}$).

*Online closed-loop performance evaluation*
While offline pseudo-online studies give an initial overview of the potential REW-MSLM decoding performance and benefits, they are not fully generalizable due to the lack of patient's feedback. Online experiment is the only solution to fully estimate the model performance. Therefore, online closed-loop experiments integrating REW-MSLM as neural signal decoder were achieved.

For each effector, virtual avatar or exoskeleton, a REW-MSLM decoder was recursively trained during 6 closed-loop experiments distributed over 2 months and was not reupdated since then. The total training time of the models for virtual avatar was 3 hours and 37 minutes with a total of 189, 194, 181 and 218 trials for the left and right hand translation and left and right hand rotation control, respectively. The calibration of 3 hours and 33 minutes was performed to train the model dedicated to the exoskeleton control with a total of 180, 184, 188 and 226 trials for the left and right hand translation and left and right hand rotation control.
The performance of the models were evaluated during 37 avatar experiments distributed over 5 to 203 days after the last model recalibration session and 15 exoskeleton experiments distributed over 0 to 167 days after the last model recalibration session. Five exoskeleton experiments conducted between the 62$^{nd}$ and 63$^{rd}$ days were excluded due to patient health issues unrelated to the study.

*Online Performance criteria*

Discrete performance was evaluated using the indicators previously presented for the offline performance: the accuracy ($acc$) and the F-score ($fsc$). In addition, the experiment performance was evaluated using the success rate (SR) [1,3,21] sets as the percentage of targets hit, and the R-ratio [21], defined as the ratio between the distance realized by the effector to reach a target and the distance from the initial position of the effector to the target location. R-ratio [21] is also named as the distance ratio [14] and is equivalent to the inverse of the individual path efficiency [2,3] of each task. Finally, we evaluated the evolution of the performance indicators across experiments. The linear fit with a 95% confidence interval was computed for each indicator to test the zero slope hypothesis and evaluate the performance stability across time. Supplementary videos (SV1, SV2 and SV3) present examples of sessions 36,106,167 days after the last model calibration using the exoskeleton.

To control for potential experimental biases, the chance level of the performance indicators was computed and the quality of the neuronal signal recorded during the experimental sessions was evaluated.

*Chance level study.*

Discrete states are not uniformly distributed, with a higher prior probability for idle and hand movements than wrist rotations (for exoskeleton-based experiments: idle, left and right hand, left and right wrist states represented 26%, 36%, 27%, 6%, 5% of the discrete state distribution, respectively). For the SR and R-ratio, $n = 100$ random hit experiments were repeated. Random movement reaching tasks were performed with the same target locations as those used during the exoskeleton-based experiments. A 3D randomly moving cursor must reach a randomly selected target within a fixed duration (defined as the 99% of the cumulative distribution of the experimental time used in the exoskeleton-based experiments). At each time step, the cursor moved in a 3D random direction with a speed fixed to the maximal speed of the exoskeleton. A target was considered a hit when the distance between the cursor and the target was less than 5 cm. These random sessions resulted in an averaged SR of $7.1 \pm 5.5$% (R-ratio: $24 \pm 14$) for the left hand translation, $9.5 \pm 6.6$% (R-ratio: $33 \pm 19$) for the right hand translation, $40 \pm 7.1$% (R-ratio: $15 \pm 4.6$) for the left hand rotation and $33 \pm 4.9$% (R-ratio: $12 \pm 2.7$) for the right hand rotation tasks.

*Neuronal signal recording quality evaluation*

The ECoG recorded at rest prior to each experiment was analyzed to assess the signal quality over the sessions performed with an avatar or in the exoskeleton. Because of recording issues, the rest sessions recorded on day = 168 and day = 167 after the last model calibration were removed for the virtual and exoskeleton sessions respectively. A 90s time window (from + 20 s to + 110 s post-recording onset) was used to calculate the power spectral density on the demeaned 64 electrodes using a 4$^{th}$ order Butterworth, IIR filter. Bandpower values (dB) were computed for the whole frequency range used in the study (10-150 Hz) and for the two frequency bands which are generally used in ECoG-driven BCI studies: 20-40 Hz and 60-110 Hz[13,18,49]. For each frequency band, the bandpower values were fitted to a linear regression to estimate the corresponding slope and its error-estimate with a 95% confidence interval.

*Model convergence evaluation*

The convergence of the models created during the online closed-loop asynchronous alternative 8D experiments using the avatar and the exoskeleton were studied. The Frobenius distance was evaluated between consecutive update of the models during the calibration period for the expert models dedicated to the 3D left and right hand translation decoding and 1D left and right wrist rotation decoding. The Frobenius distance is the generalization of the Euclidian distance applied to tensors.

## Results

Pseudo-online REW-MSLM decoding performance evaluation

*REW-MSLM discrete pseudo-online performance.*

The REW-MSLM demonstrated strong discriminative abilities (Figure 5a) between all states ($acc = 93 \pm 1.8\%$, $fsc = 86 \pm 3\%$), between IS and AS ($acc = 91 \pm 3\%$, $fsc = 84 \pm 5\%$) and between $AS_{LH}$ and $AS_{RH}$ ($acc = 99 \pm 0.8\%$, $fsc = 99 \pm 0.8\%$) regardless of the experimental paradigm. The same performance indicators lead to $acc = 87 \pm 2\%$, $fsc = 76 \pm 3\%$ between all states, $acc = 86 \pm 2\%$, $fsc = 75 \pm 3\%$ between IS and AS and $acc = 93 \pm 0.3\%$, $fsc = 93 \pm 0.2\%$ between $AS_{LH}$ and $AS_{RH}$ for REW-SLM algorithm whereas REW-NPLS performs $acc = 62 \pm 2\%$, $fsc = 36 \pm 5\%$ between all states, $acc = 70 \pm 7\%$, $fsc = 49 \pm 0.6\%$ between IS and AS and $acc = 59 \pm 8\%$, $fsc = 57 \pm 9\%$ between $AS_{LH}$ and $AS_{RH}$. The REW-MSLM strongly discriminated each state with a particularly robust distinction between the left and right hand. Significant improvements compared to REW-NPLS$_T$ and REW-SLM were evident in the majority of the performance indicators (Figure 5a). No significant differences between the performance in the experimental sessions B and C were found ($p > 0.1$), indicating model stability in session C, even though the model was not recalibrated in these experiments.

The latency of the switching state averaged over the three experimental paradigms (A, B and C) was higher for the REW-MSLM than for the REW-SLM: $lat = 2.05 \pm 0.059$ s versus $lat = 1.46 \pm 0.31$ (Figure 5b). Similarly, the error block duration increased with the REW-MSLM decoders. The HMM state decoder error lasted $ErrB_{time} = 4.31 \pm 0.88$ s, whereas the discrete static decoder error duration of the REW-SLM was $ErrB_{time} = 0.49 \pm 0.024$ s. However, the error block rate decreased considerably with the REW-MSLM decoders: the error block rate for the REW-SLM was high ($ErrB_{rate} = 20.7 \pm 1.95$ error blocks per minute), whereas that of the REW-MSLM was reduced to $ErrB_{rate} = 1.6 \pm 0.26$ blocks per minute.

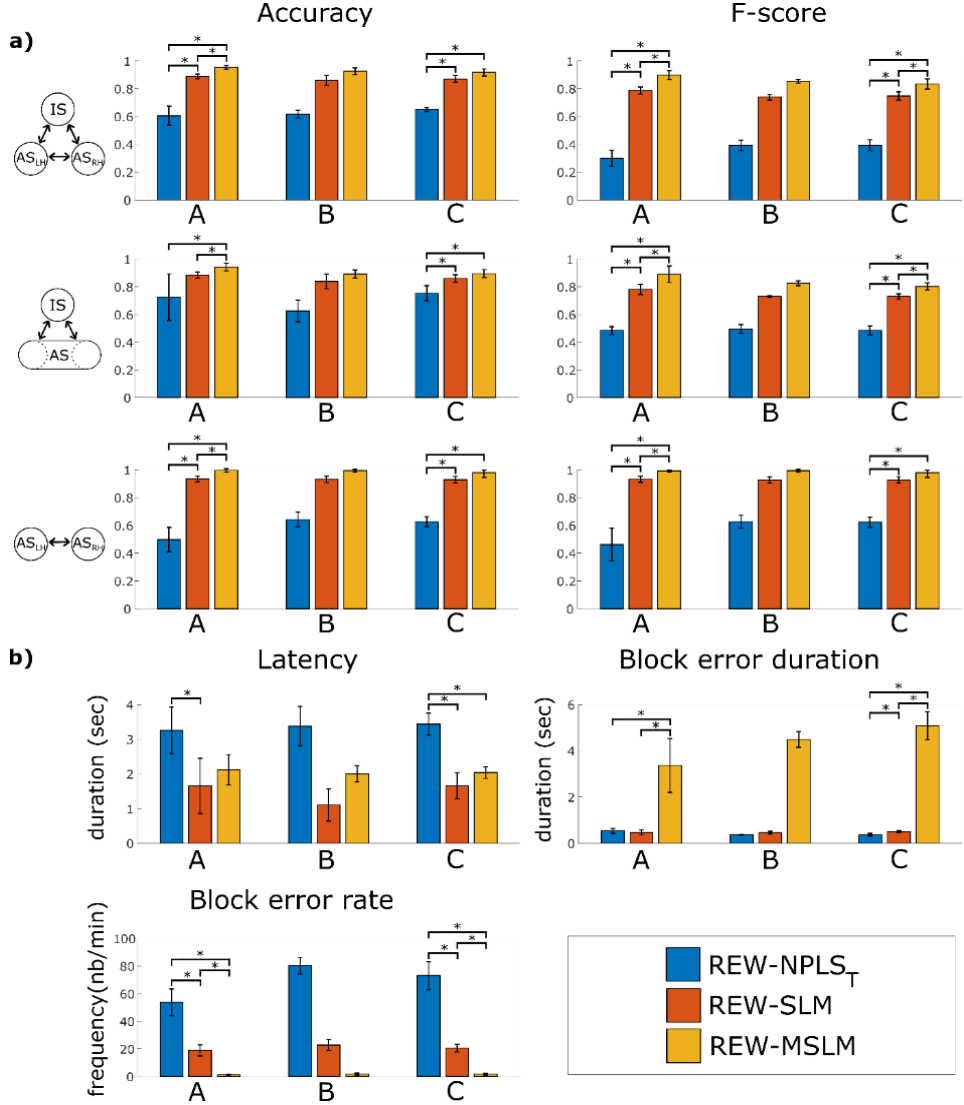

*Figure 5: **State decoding results obtained during pseudo online experiments.** a) Average accuracy and F-score over datasets A, B and C for 3 different analyses: all states (idle state IS , left hand translation active state $AS_{LH}$ and right hand translation active state $AS_{RH}$) considered independently, IS versus AS (both hand translation states merged) performance and $AS_{LH}$ versus $AS_{RH}$. b) Time dynamic performance indicators: Latency duration is evaluated as the time required to reach the desired state. Error block durations shows the average time of the consecutive misclassified samples. The error block rate represents the occurrence of blocks of wrong detections per minute. Standard deviation is represented for each algorithm and each dataset using a vertical bar. Significance of the differences between the three decoders are computed for datasets A and C (B is excluded because of the sample size) using the Mann-Whitney U test with Bonferroni corrections ($\alpha_{multi-class} = 0.0167$) in the multi-class comparisons. Otherwise, $\alpha = 0.05$. Significant values are indicated by an asterisk.*

*REW-MSLM continuous pseudo-online performance.*

To evaluate expert-specific subset training strategy piece-wise linear continuous REW-MSLM predictions were compared to those of the REW-NPLS decoder trained on the entire dataset. No statistical differences in the decoding performance were highlighted between REW-MSLM and REW-NPLS. For the paradigm A, $CosSim_{LH} = 0.095 \pm 0.05$, and $CosSim_{RH} = -0.03 \pm 0.16$ in average for the REW-MSLM decoder compared to $CosSim_{LH} = -0.03 \pm 0.14$, and $CosSim_{RH} = -0.04 \pm 0.1$ for the REW-NPLS model (Figure 6a). Left hand decoding of experimental sessions B and C demonstrated equivalent average decoding performance: $CosSim_{LH} = 0.21 \pm 0.06$ and $CosSim_{LH} = 0.23 \pm 0.13$ for experimental sessions B and C for REW-MSLM decoder and $CosSim_{LH} = 0.18 \pm 0.05$ and $CosSim_{LH} = 0.18 \pm 0.11$ for experimental sessions B and C for the REW-NPLS model. Right hand decoding average performance of REW-MSLM (B: $CosSim_{RH} = 0.15 \pm 0.07$ and C: $CosSim_{RH} = 0.2 \pm 0.03$) is similar to the decoding performance of REW-NPLS (B: $CosSim_{RH} = 0.14 \pm 0.09$ and C: $CosSim_{RH} = 0.19 \pm 0.03$) (Figure 6b and Figure 6c). Significant improvements in performance between dataset A and datasets B and C highlighted the benefits of cross-session training for increasing both the training data length and robustness to signal variability. No statistically significant performance differences were observed between datasets B and C, stressing the model robustness.

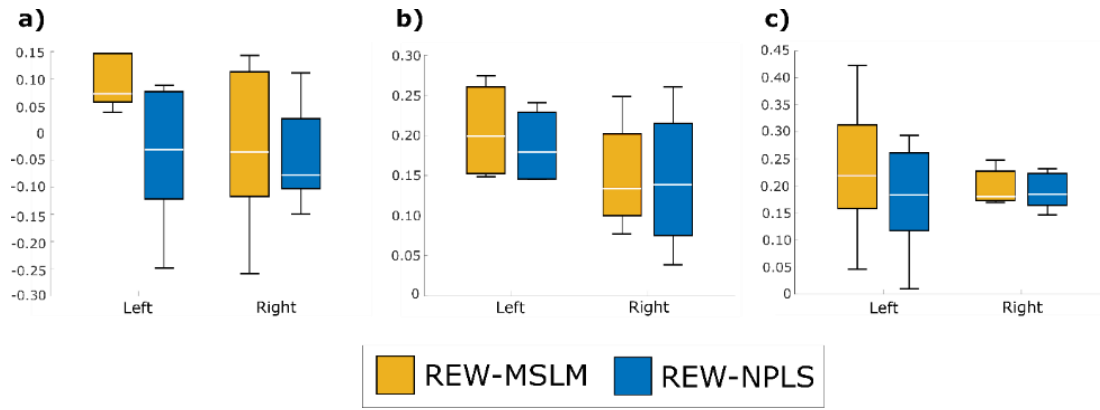

*Figure 6 **Continuous decoding performance for each hand for datasets A-C**. Statistics of the scalar product between the predicted hand directions and the optimal prediction (defined as the target-cursor oriented distance) averaged over time and the experiments of each dataset. The scalar products are represented for the left and right hand performance. The performance indicators are shown in blue for the state-of-the-art REW-NPLS model and in yellow for the new REW-MSLM. On each box, the central mark indicates the median, and the bottom and top edges of the box indicate the 25th and 75th percentiles, respectively. The whiskers extend to the extreme data.*

Online closed-loop REW-MSLM performance evaluation

*8D virtual avatar control performance*

When considering the whole frequency range used in this study (10-150 Hz), the ECoG analysis performed at rest showed a stable bandpower with a slope of -0.84% (CI = ± 0.61%) and -0.99% (CI = ± 0.84%) for the avatar and the exoskeleton experiments respectively. A similar trend was observed for the two frequency ranges which were mostly used by the decoder: the 20-40 Hz band with the respective slopes of -0.97% (CI = ± 0.58%) and -0.75% (CI = ± 0.59%) for the avatar and the exoskeleton experiments, and the 60-110 Hz band with the respective slopes of -0.23% (CI = ± 0.26%) and -0.13% (CI = ± 0.65%) for the avatar and the exoskeleton (Figure 7).

High classification decoding performance discriminating five states (idle, left and right hands translation and left and right wrists rotation) was demonstrated with the REW-MSLM algorithm across all the experiments with an average (across states and experiments) F-score of $fsc = 76 \pm 9\%$ and accuracy of $acc = 93 \pm 3\%$ (Figure 8a). The hit performance demonstrated a right hand translation SR of $53 \pm 15\%$ (R-ratio: $5.4 \pm 3.5$) and a left hand translation SR of $55 \pm 18\%$ (R-ratio: $5.2 \pm 3.1$), whereas the average wrist rotation SR was $95 \pm 8.2\%$ (R-ratio: $3.6 \pm 3.3$) across all the experiments.

The zero slope hypothesis was not rejected for 16 of the 18 indicators. It was rejected for the left wrist rotation R-ratio, which increased by $0.014$ daily, and the right hand translation SR, which reduced daily by $0.07\%$. These results highlight the stability of the REW-MSLM over 6.5 months using a virtual avatar effector during 8D experiments.

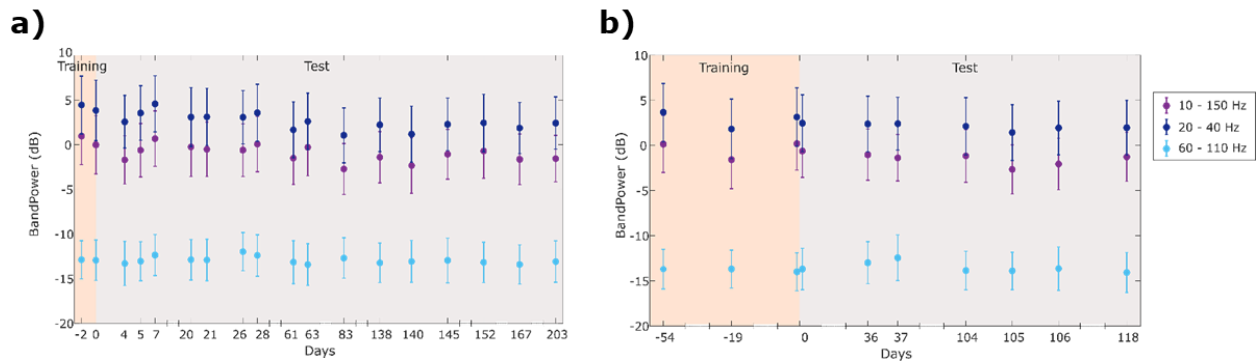

*Figure 7: Evolution of the mean bandpower values (dB ± sd) over days for the avatar (a) and the exoskeleton (b) experiments, with day = 0 being the last day the model calibration was updated. The bandpower values were computed for three frequency ranges of interest: 10-150 Hz, 20-40 Hz and 60-110 Hz.*

*8D exoskeleton control performance*

The discrete decoding performances of 8D exoskeleton control experiments yielded relevant and stable results across the 167 days. The REW-MSLM's gating yielded an average F-score of $75 \pm 12\%$ and accuracy of $92 \pm 4\%$ with high distinctiveness between the classification of the left and right sides of the body (less than 1% of misclassified samples) and strong idle state decoding with an average of 85% accurately classified idle state samples.

Left hand translation demonstrated an average SR of $69 \pm 13\%$ with an R-ratio of $6.7 \pm 5.4$. Right hand translation showed similar SR but less stable than left hand translation, with an average SR of $65 \pm 29\%$ and an R-ratio of $13 \pm 4.5$ (Figure 8b). The decoding for both wrists showed fast reaching performance with an average right and left wrist rotation task completion rate of $93 \pm 12\%$ and R-ratio ($2.9 \pm 2.4$). It is worth to note, that for the period 0 to 37 days after the last decoder calibration session, the online sessions using the exoskeleton yielded a decoding accuracy of 94% averaged across the five classes. Additionally, an average SR for both hands of 83% and 97% with an average R-ratio of $6.4 \pm 2.3$ and $3.3 \pm 1.7$ for the 3D hand translation and 1D wrist rotation were reported for 8D control on the same period. This period corresponds or overpasses the time interval reported generally in ECoG-based BCI studies. Commonly, ECoG based clinical trials last from several days to 1 or 2 weeks of research with an implantation from 3 to 35 days [10,12–15,18,23,24,51,74,75].

Decoding stability was evaluated with zero slope hypothesis, which was not rejected for 12 of the 18 indicators. The right side of the body seemed to have a slow performance decrease across experiments, gathering 5 of the 6 diminishing indicators. The linear fits demonstrated significant decreases in right limb performance for the discrete right wrist rotation indicators ($-0.25\%$ F-score and $-0.04\%$ accuracy per day) and for the right hand translation F-score ($-0.17\%$), SR ($-0.42\%$) and R-ratio ($+0.24$). Significant decreases were found in the left hand SR ($-0.18\%$ per day). The left hand SR seemed to decay in the first experiments before stabilizing.

All the 18 performance indicators had higher values than those obtained by chance level studies for all the experiments. Chance level studies highlighted an averaged SR of $7.1 \pm 5.5\%$ (R-ratio: $24 \pm 14$) and $9.5 \pm 6.6\%$ (R-ratio: $33 \pm 19$) for the left and right hand translation respectively whereas left and right hand rotation tasks chance level was evaluated at, $SR = 40 \pm 7.1\%$ (R-ratio: $15 \pm 4.6$) and $SR = 33 \pm 4.9\%$ (R-ratio: $12 \pm 2.7$).

Figure 9 illustrates the convergence through the model update iterations during 6 calibration session of coefficients of the expert models for the left and right hand translation and left and right wrist rotation decoding.

Examples of hand trajectories performed on the session 106 days after the model calibration are presented in Figure 10a and Figure 10b for the left and right hand translation, respectively. Additional trajectories are proposed in the Supplementary Materials (Figure S1) and supplementary videos (SV1, SV2 and SV3) present examples of sessions 36, 106, 167 days after the last model calibration.

The entire session of the 106th days is represented in Figure 10c. This session is composed of successive tasks with a total of two right hand translation tasks and three idle, left hand translation, left and right hand rotation tasks. Each tasks is composed of several trials. Trajectories from Figure 10a and Figure 10b are trials form the first left hand and second right hand translation tasks. Gating model used for exoskeleton control is represented on the spatial, frequency and temporal modality in Figure 11. Spatial modality presents heavy parameter weights on the contralateral electrode array for left and right hand (translation and rotation) states. Both translation tasks present similar model with dominant frequency band between 20-30Hz (β-band) and 80Hz-120Hz (γ-band). The same frequency band are relevant for rotation and idle state model, nevertheless, lower frequency band (<20 Hz) significantly contribute to the decoding, especially for idle state decoding. Parameter weights in the temporal modalities are similar for all states, emphasizing parameters between 0.5s and 0.1s. Variability of model coefficients according to the different modalities are presented in the supplementary materials in Figure S2.

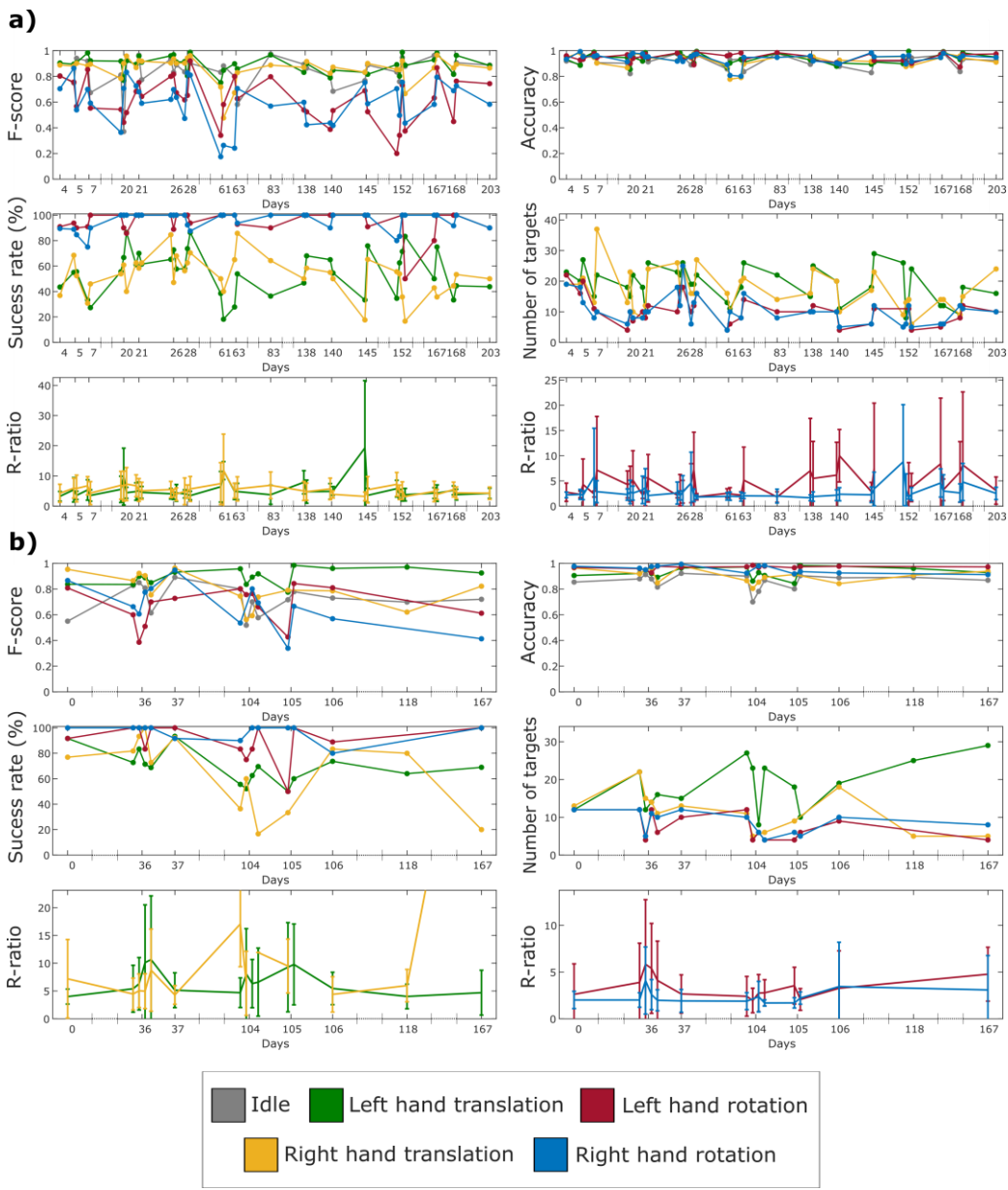

*Figure 8: **Online experiment performance across several months for the virtual avatar and exoskeleton effectors.** Online 8 DoF experiment performance (for each state: idle, left and right hand translation and rotation)*

*using virtual avatar effector across 203 days after last model calibration (a) or using exoskeleton effector across 167 days after last model calibration (b). F-score and accuracy discrete performance indicators are evaluated for each state. Continuous performances are computed using the success rate (SR) (percentage of targets hit) and the R-ratio (ratio between the distance travelled by the effector to reach a target and the distance from the initial position of the effector to target location). Standard deviation is represented for each algorithm and each dataset using a vertical bar.*

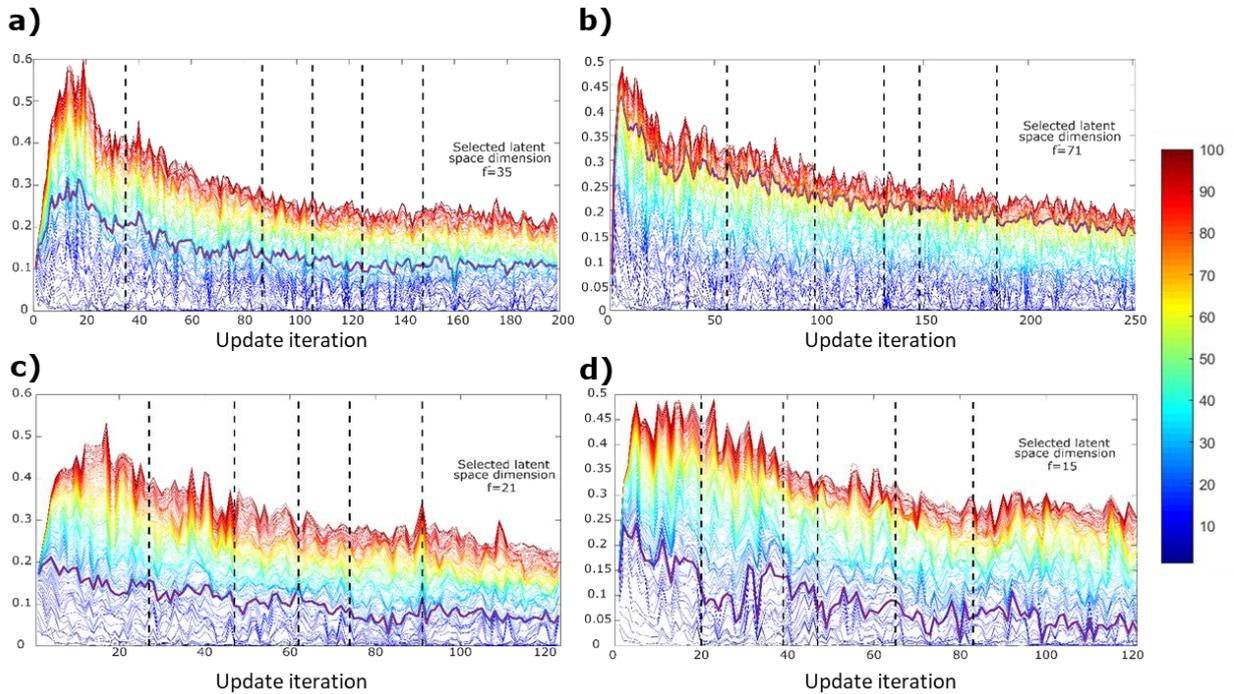

*Figure 9 Convergence through the model update iterations of coefficients of the expert decoding models of left hand translation (a), right hand translation (b), left wrist rotation (c), right wrist rotation (d). The Frobenius distance between consecutive coefficients update is depicted by coloured lines for a set of updated models including $f$ factors, $f = 1, ... , 100$. Models selected by online validation procedure are depicted by the bold purple line. Calibration sessions are separated by the black vertical dotted lines.*

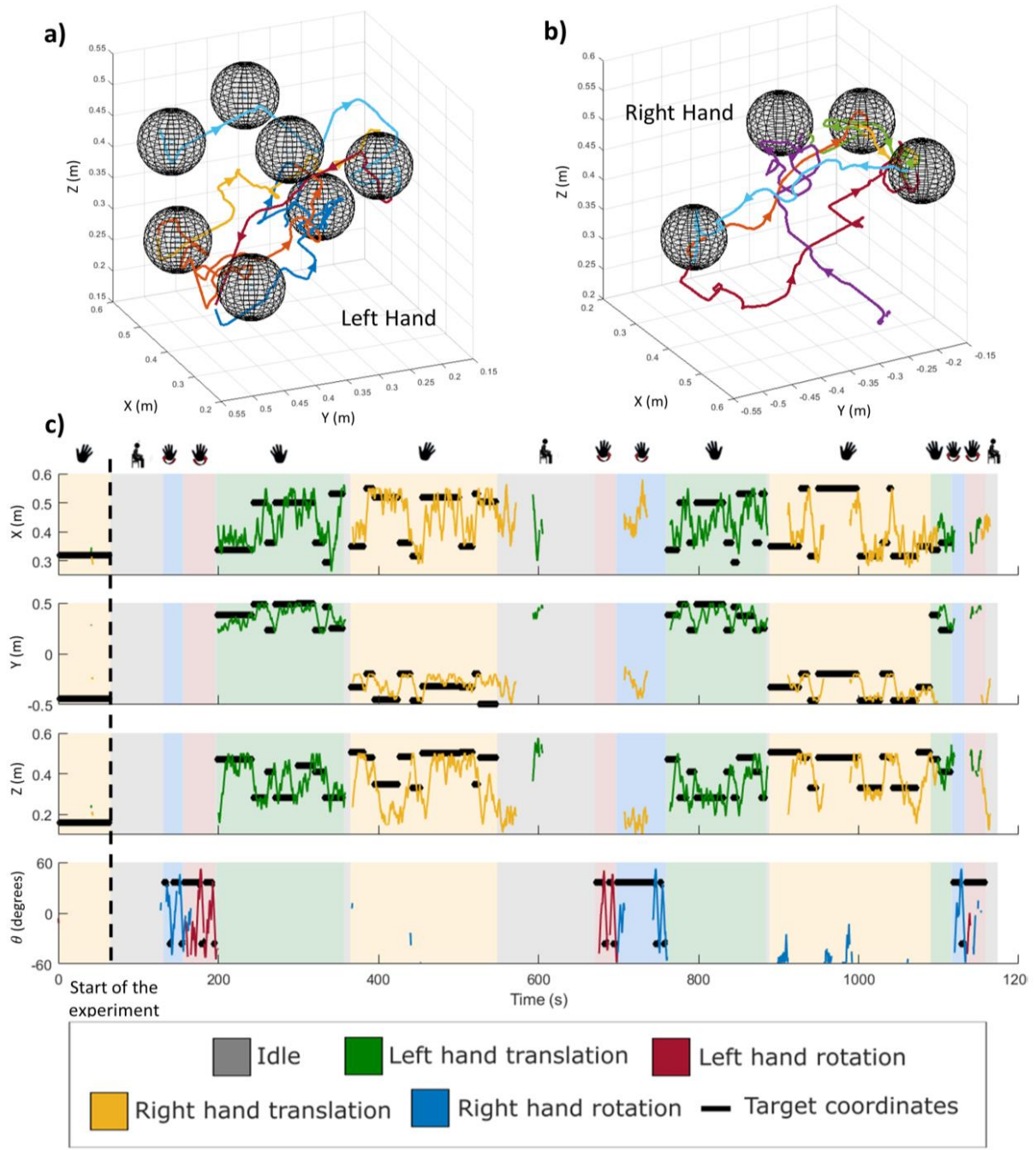

*Figure 10:* **Session realized 106 days after the last model calibration using exoskeleton effector**. A) left hand trajectory across time and trials. The trajectories are extracted from the first left hand task of the session. B) right hand trajectory across time and trials. The trajectories are extracted from the second right hand task of the session. Each color represents one trial, the trajectory to reach one specific target. C) Movement on X, Y, Z and θ (angle for wrist rotation) across the sessions performed 106 days after the last model calibration. Shaded area color correspond to the task that patient must perform. Colored Lines represent left and right hand coordinates for X, Y and Z-axis and left and right wrist angle for θ axis. Other examples of left and right 3D hand translation trajectories are available on the Supplementary Materials

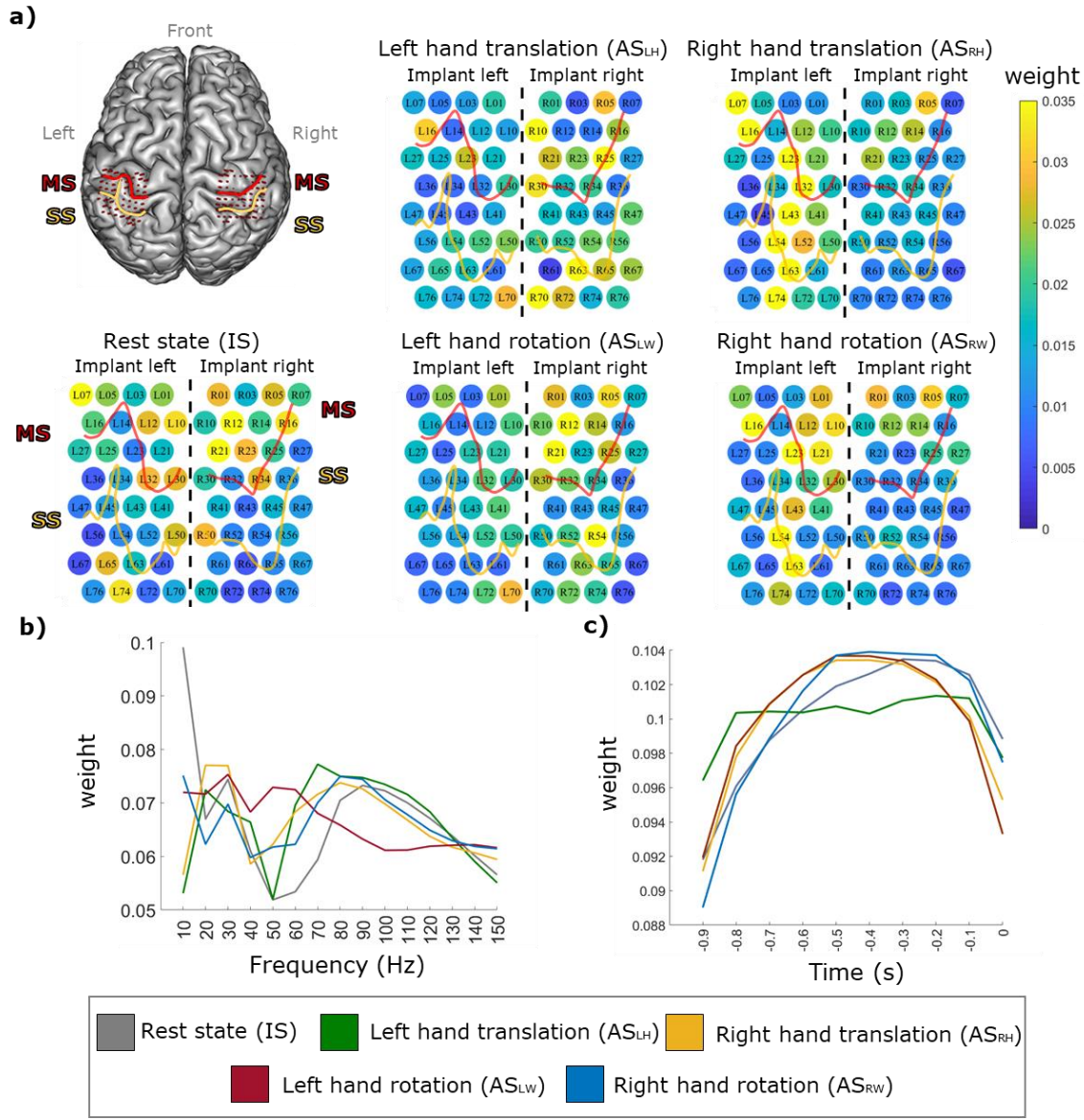

*Figure 11: **Example of a gating model.** Gating parameter weights (discrete decoding) of the REW-MSLM created using an exoskeleton effector according to the spatial (A), frequency (B) or temporal (C) modalities for each state: rest state (IS), left hand 3D translation and rotation state ($AS_{LH}$ and $AS_{LW}$) or right hand 3D translation and rotation state ($AS_{RH}$ and $AS_{RW}$). The sensory sulcus (SS) and motor sulcus (MS) are represented in the spatial domain in yellow and red curves respectively. The spatial modality shows, as expected, strong parameter weights on the contralateral electrode array for left and right hand (translation and rotation) states; in addition, translation and rotation from the same hand seem to activate nearby but distinct electrodes. The frequency modality highlights the beta and high gamma frequency bands as relevant frequencies for state decoding, whereas the temporal modality emphasis parameters from 0.5s to 0.1s. The variability of the state decoding model coefficients according to the time and frequency modalities is represented in the supplementary materials (Figure S2).*

**Discussion**
Based on the first successful long-term (more than 36 months) chronic exploitation of bilateral epidural ECoG recordings in a tetraplegic individual [21], we developed the REW-MSLM decoder to address the poorly explored field of asynchronous multi-limb effector control. This decoder was designed to overcome the major issues related to the translation of BCIs from the laboratory to real-life applications, such as the high-dimensional control of effectors based on chronic neural recordings, experiments closer to real life behaviour, and the ability of the BCI system to act as a stand-alone device and switch between IS and AS phases. ME architecture was employed to handle numerous dimensions and to decode the robust idle state. To allow cross-session training of the decoder with multiple recording conditions during closed loop BCI effectors control experiments, we developed an adaptive/incremental learning algorithm that handles high-dimensional tensor data flow. Tensor-based algorithms emerged as a promising strategy for brain signal processing allowing simultaneous treatments of high-dimensional data in the temporal, frequency and spatial domains [19,66]. Dynamic expert gating using a HMM was added to ME decoder to ensure the robustness of states. The proposed algorithm is fully adaptive. It learns the model in an incremental manner, including the hyperparameters.

The REW-MSLM was integrated into a custom-made BCI adaptive brain signal decoder (ABSD)[21] software platform to provide a tetraplegic patient with the control of a virtual avatar and an exoskeleton in real time. Volitional alternating rotation and 3D translation movements of both hands could be executed. This performance was achieved using EpiCoG recordings which are less invasive than the subdural ECoG recordings reported in most of the BCI studies.

ME structure benefits for multi-limb effectors control
The REW-MSLM architecture fits the multi-limb paradigm. Each expert can be associated to a particular limb or action while the HMM gating model aims to establish the state selection and handle robust idle state detection for complex asynchronous state decoding [17]. Moreover, we hypothesized, based on well-established neurophysiological knowledge [20,31], that neural data associated with each limb or specific action can be partitioned into different neural regions/patterns. Consequently, each expert was only trained on a small subset of the entire training dataset. Such training allowed individual update of the experts, and incrementally appending new experts to control new dimensions without full re-training of other experts. To demonstrate the relevance of the ME model structure, and the importance of dynamic vs. static gating, we compared the REW-MSLM to the state-of-the-art adaptive algorithms.

The comparison of several algorithms is a conventional tool to conclude on the improvements obtained with the proposed algorithms. However, comparing several online algorithms during closed-loop experiments is a complicated task as, during such experiments, the predicted trajectories are related to the current decoding model and patient's feedback. Consequently, it is not possible to compare in online closed-loop experiments several algorithms that produce different predictions and feedbacks. Several series of online closed loop sessions are particularly time consuming. In the current study, offline comparison study was undertaken in pseudo-online manner with 3 databases. The datasets were recorded using 3 different experimental paradigms (single session decoder training, cross session decoder training, fixed decoder) during online closed-loop experiments using conventional REW-NPLS algorithm as the decoder. While offline studies gave an initial overview of the potential REW-MSLM decoding performance and benefits, they were not generalizable due to lack of the appropriate user feedback. Nevertheless, it allowed characterizing the studied algorithms before an integration into the clinical BCI decoding platform.

For discrete decoding, the REW-MSLM outperformed alternative approaches in discrete classification regardless of the dataset and paradigm with an averaged F-score improvement across all paradigms of $39 \pm 4\%$ and $8.3 \pm 2\%$ compared to the REW-NPLS and the REW-SLM respectively. These results confirmed the benefits to train a specific model dedicated to state classification and the improvements related to dynamic classification. The switching state latency study related to the state transition delay between the instruction and the discrete decoding response demonstrated an average increase in duration by 0.45 s, 0.87 s and 0.38 s (over 3 datasets) between the discrete decoder with and without dynamic HMM processing. However, the REW-MSLM results show a drastic 92% decrease in the error block rate between the discrete decoder with and without dynamic HMM processing, overcoming the high frequency misclassified sample issue of static classifier. For physical effectors, such as an exoskeleton, which are in direct contact with the patient and has a latency of mechanical activation/deactivation of up to a few seconds, false activation should remain exceptionally rare events.

For continuous control, REW-MSLM experts highlighted slight improvement or similar performance compared to REW-NPLS, whereas the training datasets were different. REW-MSLM allows experts training using independent data sets. This may be highly profitable for progressive BCI decoder training increasing the tasks complexity.

In addition, the developed REW-MSLM and the state of the art REW-NPLS algorithms demonstrated similar decoding performance. However, numerous non-desired movements of the other limbs are observed using REW-NPLS. Unintended movements of a limb that the patient does not want to move impede the control of complex effectors such as an exoskeleton, and especially the asynchronous control with an idle state to decode. In contrast, REW-MSLM demonstrated similar decoding performance for the limb to be activated without unintentional movements from the other limbs thanks to accurate state classification. The suppression of the unintended movements leads to better visual feedbacks and concentration of the patient which may induce better model calibration.

EpiCoG based neural decoder for complex tasks completion
The control tasks, proposed to the patient during the experiments are more challenging than the usual state-of-the-art control tasks. Center-out tasks require to go from the centre of a workspace to one of the targets localized at equal distances. Moreover, after each trial (succeeded our failed), the position is reset to the initial position after few seconds of rest. In the point-to-point pursuit task experimental paradigm reported in this article, the patient controls the effector all along the session and without resetting the hand position. This control task is more complex because the initial position of the hand changes constantly, and decoding mistake/drifting of the hand from one trial affect the following trial. A point-to-point pursuit task is more complex compared to conventional center-out tasks in terms of explored space due to multiple (arbitrary under the constraints of control region) possible starting points and numerous targets. In the current study, 22 target positions, 11 for each hand, are proposed to the user, while a majority of center-out experiments consider 8 equally distanced targets [13,14,18,76]. Point-to-point pursuit tasks are more representative of daily life applications, and have less restricted experimental conditions.
In addition, asynchronous and alternative bimanual point-to-point pursuit experiments support rest period as well as asynchronous switch between active control tasks without external intervention. All dimensions of control (8 in general) are available to the user at any moment. While not all degrees of freedom may be active simultaneously, any point in the control region (8D) may be achieved by user at his own intention.

In the beginning of the experiments, the patient optimized the motor imagery strategy to allow controlling the effectors. He reported that he was able after several months of training to control the effectors unconsciously, without focusing on motor imagery.

Closed-loop decoder stability using EpiCoG recordings
Generally ECoG-based BCI studies are performed using temporary ECoG subdural grid implantation from 3 to 35 days post-surgery [10,12–15,18,23,24,51,74,75]. In our experiments, the online closed-loop SR for both effectors realized from 0 to 37 days after the last model calibration (Figure 8) are similar or above current ECoG-based state of the art performance for 3D decoding. Importantly, compared to these subdural ECoG studies we did not perform any model recalibration during this time period even though we used a system which is less invasive [14].

The online closed-loop results presented a high stability level and were far above the realized chance level study across all experiments for both effectors. For the exoskeleton experiments, the left hand translation SR seemed to decay between the 37th and the 104th day and stabilize until the end, whereas the right hand translation SR showed higher variability in the performance (between 17% and 100%). For discrete decoding, switching from left arm control to right arm control (and vice versa) produced less than 1% of the errors. Most of the decoding misclassifications were related to two issues. First, the majority of the mistakes were related to false positive idle state activation. Second, the decoders struggled to differentiate between rotation and translation from the same limb. These difficulties may be related to the similarity of both tasks and may consequently lead to brain neural signal pattern activations within close areas.

Our results seem to demonstrate higher average performance to control the exoskeleton than the virtual avatar. This could be explained by the fact that the exoskeleton provides a more realistic feedback to the patient than the virtual avatar. However, it is difficult to make any conclusion due to the small number of experiments considered in this study.

The online control of both effectors was maintained, without recalibration, over 6 months of clinical experiments (for 167 days and during 203 days for the exoskeleton and virtual avatar effectors, respectively), indicating the stability of both the REW-MSLM decoder and the neural activity recording method with the two WIMAGINE EpiCoG recording implants[53]. These results show that this system outperforms the state-of-the-art ECoG-based BCIs, and outperforms both the state-of-the-art ECoG and MEAs-based BCIs in terms of decoder stability.

The pseudo online study induces the benefits of cross-session training for obtaining a better decoder, more robust to brain and experimental condition variability. Indeed, continuous performance was low for dataset A (single session decoder training). Results from dataset C (fixed decoder) showed stable performance whereas the model was trained on the basis of cross-session calibration procedure from dataset B recorded 9 to 28 days before. In the online study, the REW-MSLM was trained for each effector based on cross-session calibration procedure for 6 experiments over 6 days, distributed over 2 months, for approximately 3.5 h (with in averaged 195 trials). The duration of the model training periods seems moderate, considering the high number of dimensions to control and performance obtained compared to those in similar studies [13,14]. More training data may lead to a more generalized model and thus, better results.

Figure 11 illustrates gating model weights in the frequency, temporal and spatial modalities. In the frequency modality the model coefficients are consistent with the previous studies which highlighted the significance of β and high γ-band to decode movements from direct neural signals[18,49,74]. As expected, spatial weights were higher in the contralateral electrodes of the realized movement for both left and right hand translation and rotation which is corroborated by previous studies[49,78,79].

Limitations and perspectives
The current paper reports the long-term stability of high dimensional (8D) control of bimanual exoskeleton and its avatar. While the study demonstrates promising results, they were demonstrated for a single patient. The implantation of 3 more patients is planned in the "BCI and tetraplegia" clinical trial protocol and would provide more data to support the conclusions of this article.

Offline comparative study was undertaken to evaluate the proposed algorithm benefits. Although offline studies give an initial overview of the potential decoding performance and benefits, they are not generalizable to the case of human-in-the-loop systems due to lack of appropriate feedback in the data.

A restricted dataset was used for decoders training. Decoding models were fixed without determining an optimal training time. More training data may lead to a more generalized model and better results. The optimization of training paradigm is one of the perspectives of the presented research. The model will be trained for a longer time to accumulate more information and evaluate the impact of a larger dataset on decoding performance. Model interpretation and convergence will be further investigated. In addition, patients' adaptations and improvements will be analyzed to evaluate the impact of experiment frequency on performance stability

Only alternative bimanual control was performed due to experimental paradigm. However, simultaneous bimanual control is theoretically possible thanks to REW-MSLM soft gating strategy: the gating is not a selection of one limb among the others but the mixing of all of them depending on the probability of limb activation computed by the HMM gating. Simultaneous bimanual effector control is a nearest perspective of the study.

The REW-MSLM benefits from a mixture of experts architecture, which splits the dataset to train particular experts. Continuous decoders are responsible for a single or group of dimensions. This structure allows us mixing experts from different training sets and different models or adding new dimensions without retraining all the experts. Increasing control complexity by adding dimensions sequentially is highly profitable for patient training. The mixture of experts decoder architecture favors further increase of control dimensions. Doubling the resolution of the recording system is expected in near future and may allow an increase in number of degrees of freedom.

Compared to traditional center-out tasks, the current study reports an experimental paradigm less restrictive in term of experimental conditions, with a wider exploration of the control space. Experiments closer to domestic, urban, and professional environments should be designed to move the technology from clinical trials to daily life applications.

Faster and straighter reaching trajectories are likely to be particularly profitable for patients. Various post-processing strategies will be investigated in future studies to provide better control and feedback to the patient. As the drop of decoding performance in the target neighborhood is regularly observed in BCI studies, alternative ME architecture with states associated to movement phases will be explored [68].

**Appendix**

**Performance Indicators.**
**Discrete performance Indicators.** We evaluated discrete performance using accuracy ($acc$) and F-score ($fsc$) indicators. These indicators are computed using the confusion matrix, which summarizes the number of correctly classified samples from one state (true positives, $tp$), incorrectly labelled samples in one state (false negatives, $fn$), correctly classified samples not belonging to the state (true negatives, $tn$) and incorrectly labelled samples not belonging to the state (false positives, $fp$):

$$Accuracy = \frac{1}{K}\sum_{k=1}^{K} \frac{tp_k + tn_k}{tp_k + tn_k + fp_k + fn_k},$$

$$Fscore = \frac{1}{K} \cdot \sum_{k=1}^{K} \frac{(\beta^2 + 1)\, Precision_k\, Recall_k}{\beta^2\, Precision_k + Recall_k},$$

$$Precision_k = \frac{tp_k}{tp_k + fp_k},\ Recall_k = \frac{tp_k}{tp_k + fn_k}.$$

The weighting coefficient $\beta$ was set to one, the true positives $tp_k$ are considered for samples labelled as belonging to state $k$, and the true negatives $tn_k$ include those from all the other states (one versus all analysis). $K = 3$ for the pseudo online comparison study, and $K = 5$ for the 8 D online experiments. Accuracy is the indicator generally used in BCI for binary and multi-state classification [18,32,57,80,81] and is useful for performance comparison due to its ease of computation and interpretation. Nevertheless, as accuracy presents weaknesses in the case of highly unbalanced class, F1-score is also computed.

The previously described state decoding indicators are sample-based performance estimators. They do not reflect the dynamic behaviour of the misclassified samples. Therefore, supplementary indicators were introduced. First, the latency between the instruction and estimated state transition was computed to evaluate the combined response time of the patient and the model. The estimated state transition was considered valid only when the decoded state was stable for 1s (10 samples). The transition must be achieved in the 5s following the instruction state transition for it to not be counted as an incorrectly labelled state. Samples belonging to the transition/latency period were not considered in the other discrete performance indicators. Finally, the blocks of consecutive misclassified samples were counted to evaluate the error block rate and their averaged durations.

**Continuous variable decoding indicator.** As mentioned above, the trajectories performed during the online closed-loop experiments are related to the decoding model used during the experiments and patient's feedback. A sample-based indicator is applied to compare the predictions of several algorithms.

The indicator $CosSim$ is based on the comparison between the predicted $\hat{\mathbf{y}}_t$ and the optimal prediction $\mathbf{y}_t$ defined as the 3D Cartesian vector between the current position and the target $\mathbf{y}_t$ for 3D translation tasks. After normalization and averaging of the scalar product

$$CosSim = \frac{1}{T}\sum_{t=1}^{T}\frac{\mathbf{y}_t \cdot \hat{\mathbf{y}}_t}{\|\mathbf{y}_t\|\|\hat{\mathbf{y}}_t\|},$$

where "·" defined the dot product, and $T$ is the number of samples recorded for a specific limb (right or left hand). CosSim indicator varies from -1 to 1 evaluating the algorithm's global static prediction performance.

**Online performance Indicators.** Online performance is evaluated using the success rate (SR) [1,3], which is the percentage of targets hit, and the R-ratio is defined as the normalized path length of the reach. The R- ratio is the ratio between the distance travelled by the effector to reach a target and the optimal distance from the initial position of the effector to the target location. R-ratio [21] is also named distance ratio [14] and is equivalent to the inverse of the individual path efficiency [2,3] of each task. The SR and R-ratio performance indicators are defined in the same way for the evaluation of wrist rotation performance.

**References**

1. Hochberg, L. R. et al. Reach and grasp by people with tetraplegia using a neurally controlled robotic arm. *Nature* **485**, 372–375 (2012).
2. Collinger, J. L. et al. High-performance neuroprosthetic control by an individual with tetraplegia. *The Lancet* **381**, 557–564 (2013).
3. Wodlinger, B. et al. Ten-dimensional anthropomorphic arm control in a human brain–machine interface: difficulties, solutions, and limitations. *J. Neural Eng.* **12**, 016011 (2015).
4. Murphy, M. D., Guggenmos, D. J., Bundy, D. T. & Nudo, R. J. Current Challenges Facing the Translation of Brain Computer Interfaces from Preclinical Trials to Use in Human Patients. *Front. Cell. Neurosci.* **9**, (2016).
5. Ward, M. P., Rajdev, P., Ellison, C. & Irazoqui, P. P. Toward a comparison of microelectrodes for acute and chronic recordings. *Brain Res.* **1282**, 183--200 (2009).
6. Perge, J. A. et al. Intra-day signal instabilities affect decoding performance in an intracortical neural interface system. *J. Neural Eng.* **10**, 036004 (2013).
7. Sussillo, D., Stavisky, S. D., Kao, J. C., Ryu, S. I. & Shenoy, K. V. Making brain–machine interfaces robust to future neural variability. *Nat. Commun.* **7**, 13749 (2016).
8. Lebedev, M. A. & Nicolelis, M. A. L. Brain-Machine Interfaces: From Basic Science to Neuroprostheses and Neurorehabilitation. *Physiol. Rev.* **97**, 767–837 (2017).
9. Rak, R. J., Kołodziej, M. & Majkowski, A. Brain-computer interface as measurement and control system The review paper. *Metrol. Meas. Syst.* **Vol. 19**, 427–444 (2012).
10. Leuthardt, E. C., Schalk, G., Wolpaw, J. R., Ojemann, J. G. & Moran, D. W. A brain–computer interface using electrocorticographic signals in humans. *J. Neural Eng.* **1**, 63 (2004).
11. Leuthardt, E. C., Miller, K. J., Schalk, G., Rao, R. P. N. & Ojemann, J. G. Electrocorticography-based brain computer Interface-the seattle experience. *IEEE Trans. Neural Syst. Rehabil. Eng.* **14**, 194–198 (2006).
12. Schalk, G. & Leuthardt, E. C. Brain-Computer Interfaces Using Electrocorticographic Signals. *IEEE Rev. Biomed. Eng.* **4**, 140–154 (2011).
13. Wang, W. et al. An Electrocorticographic Brain Interface in an Individual with Tetraplegia. *PLOS ONE* **8**, e55344 (2013).
14. Degenhart, A. D. et al. Remapping cortical modulation for electrocorticographic brain–computer interfaces: a somatotopy-based approach in individuals with upper-limb paralysis. *J. Neural Eng.* **15**, 026021 (2018).
15. Schalk, G. et al. Two-dimensional movement control using electrocorticographic signals in humans. *J. Neural Eng.* **5**, 75 (2008).
16. Shimoda, K., Nagasaka, Y., Chao, Z. C. & Fujii, N. Decoding continuous three-dimensional hand trajectories from epidural electrocorticographic signals in Japanese macaques. *J. Neural Eng.* **9**, 036015 (2012).
17. Schaeffer, M.-C. & Aksenova, T. Switching Markov decoders for asynchronous trajectory reconstruction from ECoG signals in monkeys for BCI applications. *J. Physiol.-Paris* **110**, 348–360 (2016).
18. Bundy, D. T., Pahwa, M., Szrama, N. & Leuthardt, E. C. Decoding three-dimensional reaching movements using electrocorticographic signals in humans. *J. Neural Eng.* **13**, 026021 (2016).
19. Eliseyev, A. et al. Recursive Exponentially Weighted N-way Partial Least Squares Regression with Recursive-Validation of Hyper-Parameters in Brain-Computer Interface Applications. *Sci. Rep.* **7**, 16281 (2017).
20. Choi, H. et al. Improved prediction of bimanual movements by a two-staged (effector-then-trajectory) decoder with epidural ECoG in nonhuman primates. *J. Neural Eng.* **15**, 016011 (2018).
21. Benabid, A. L. et al. An exoskeleton controlled by an epidural wireless brain–machine interface in a tetraplegic patient: a proof-of-concept demonstration. *Lancet Neurol.* **0**, (2019).
22. Chao, Z. C., Nagasaka, Y. & Fujii, N. Long-term asynchronous decoding of arm motion using electrocorticographic signals in monkey. *Front. Neuroengineering* **3**, (2010).
23. Yanagisawa, T. et al. Electrocorticographic control of a prosthetic arm in paralyzed patients. *Ann. Neurol.* **71**, 353–361 (2012).
24. Nakanishi, Y. et al. Prediction of Three-Dimensional Arm Trajectories Based on ECoG Signals Recorded from Human Sensorimotor Cortex. *PLOS ONE* **8**, e72085 (2013).
25. Nurse, E. S. et al. Consistency of Long-Term Subdural Electrocorticography in Humans. *IEEE Trans. Biomed. Eng.* **65**, 344–352 (2018).

26. Sauter-Starace, F. *et al.* Long-Term Sheep Implantation of WIMAGINE, a Wireless 64-Channel Electrocorticogram Recorder. *Front. Neurosci.* **13**, (2019).
27. Pels, E. G. M. *et al.* Stability of a chronic implanted brain-computer interface in late-stage amyotrophic lateral sclerosis. *Clin. Neurophysiol.* **130**, 1798–1803 (2019).
28. Vansteensel, M. J. *et al.* Fully Implanted Brain–Computer Interface in a Locked-In Patient with ALS. *N. Engl. J. Med.* **375**, 2060–2066 (2016).
29. Williams, J. J., Rouse, A. G., Thongpang, S., Williams, J. C. & Moran, D. W. Differentiating closed-loop cortical intention from rest: building an asynchronous electrocorticographic BCI. *J. Neural Eng.* **10**, 046001 (2013).
30. Müller-Putz, G. R., Scherer, R., Pfurtscheller, G. & Rupp, R. Brain-computer interfaces for control of neuroprostheses: from synchronous to asynchronous mode of operation / Brain-Computer Interfaces zur Steuerung von Neuroprothesen: von der synchronen zur asynchronen Funktionsweise. *Biomed. Tech.* **51**, 57–63 (2006).
31. Ifft, P. J., Shokur, S., Li, Z., Lebedev, M. A. & Nicolelis, M. A. L. A Brain-Machine Interface Enables Bimanual Arm Movements in Monkeys. *Sci. Transl. Med.* **5**, 210ra154-210ra154 (2013).
32. Hotson, G. *et al.* Individual finger control of a modular prosthetic limb using high-density electrocorticography in a human subject. *J. Neural Eng.* **13**, 026017 (2016).
33. Elgharabawy, A. & Wahed, M. A. Decoding of finger movement using kinematic model classification and regression model switching. in *2016 8th Cairo International Biomedical Engineering Conference (CIBEC)* 84–89 (2016). doi:10.1109/CIBEC.2016.7836126.
34. Flamary, R. & Rakotomamonjy, A. Decoding Finger Movements from ECoG Signals Using Switching Linear Models. *Front. Neurosci.* **6**, (2012).
35. Lebedev, M. A. & Nicolelis, M. A. L. Brain–machine interfaces: past, present and future. *Trends Neurosci.* **29**, 536–546 (2006).
36. Orsborn, A. L. *et al.* Closed-Loop Decoder Adaptation Shapes Neural Plasticity for Skillful Neuroprosthetic Control. *Neuron* **82**, 1380–1393 (2014).
37. Jarosiewicz, B. *et al.* Advantages of closed-loop calibration in intracortical brain–computer interfaces for people with tetraplegia. *J. Neural Eng.* **10**, 046012 (2013).
38. Dangi, S. *et al.* Continuous Closed-Loop Decoder Adaptation with a Recursive Maximum Likelihood Algorithm Allows for Rapid Performance Acquisition in Brain-Machine Interfaces. *Neural Comput.* **26**, 1811–1839 (2014).
39. Brandman, D. M. *et al.* Rapid calibration of an intracortical brain–computer interface for people with tetraplegia. *J. Neural Eng.* **15**, 026007 (2018).
40. Shanechi, M. M. *et al.* Rapid control and feedback rates enhance neuroprosthetic control. *Nat. Commun.* **8**, 13825 (2017).
41. Li, Z., O'Doherty, J. E., Lebedev, M. A. & Nicolelis, M. A. L. Adaptive Decoding for Brain-Machine Interfaces Through Bayesian Parameter Updates. *Neural Comput.* **23**, 3162–3204 (2011).
42. Gilja, V. *et al.* A high-performance neural prosthesis enabled by control algorithm design. *Nat. Neurosci.* **15**, 1752 (2012).
43. Vidaurre, C., Kawanabe, M., Bünau, P. von, Blankertz, B. & Müller, K. R. Toward Unsupervised Adaptation of LDA for Brain–Computer Interfaces. *IEEE Trans. Biomed. Eng.* **58**, 587–597 (2011).
44. Nicolas-Alonso, L. F., Corralejo, R., Gomez-Pilar, J., Álvarez, D. & Hornero, R. Adaptive semi-supervised classification to reduce intersession non-stationarity in multiclass motor imagery-based brain–computer interfaces. *Neurocomputing* **159**, 186–196 (2015).
45. Lotte, F. *et al.* A review of classification algorithms for EEG-based brain–computer interfaces: a 10 year update. *J. Neural Eng.* **15**, 031005 (2018).
46. Hazrati, M. Kh. & Erfanian, A. An online EEG-based brain–computer interface for controlling hand grasp using an adaptive probabilistic neural network. *Med. Eng. Phys.* **32**, 730–739 (2010).
47. Rong, H., Li, C., Bao, R. & Chen, B. Incremental Adaptive EEG Classification of Motor Imagery-based BCI. in *2018 International Joint Conference on Neural Networks (IJCNN)* 1–7 (2018). doi:10.1109/IJCNN.2018.8489283.
48. Milekovic, T. *et al.* Stable long-term BCI-enabled communication in ALS and locked-in syndrome using LFP signals. *J. Neurophysiol.* **120**, 343–360 (2018).
49. Waldert, S. *et al.* A review on directional information in neural signals for brain-machine interfaces. *J. Physiol.-Paris* **103**, 244–254 (2009).
50. Jeunet, C., N'Kaoua, B. & Lotte, F. Chapter 1 - Advances in user-training for mental-imagery-based BCI control: Psychological and cognitive factors and their neural correlates. in *Progress in Brain Research* (ed. Coyle, D.) vol. 228 3–35 (Elsevier, 2016).
51. Schalk, G. *et al.* Decoding two-dimensional movement trajectories using electrocorticographic signals in humans. *J. Neural Eng.* **4**, 264–275 (2007).
52. Ball, T., Schulze-Bonhage, A., Aertsen, A. & Mehring, C. Differential representation of arm movement direction in relation to cortical anatomy and function. *J. Neural Eng.* **6**, 016006 (2009).
53. Mestais, C. S. *et al.* WIMAGINE: Wireless 64-Channel ECoG Recording Implant for Long Term Clinical Applications. *IEEE Trans. Neural Syst. Rehabil. Eng.* **23**, 10–21 (2015).
54. Yuksel, S. E., Wilson, J. N. & Gader, P. D. Twenty Years of Mixture of Experts. *IEEE Trans. Neural Netw. Learn. Syst.* **23**, 1177–1193 (2012).
55. Chen, C. *et al.* Prediction of Hand Trajectory from Electrocorticography Signals in Primary Motor Cortex. *PLOS ONE* **8**, e83534 (2013).
56. Eliseyev, A. & Aksenova, T. Stable and artifact-resistant decoding of 3D hand trajectories from ECoG signals using the generalized additive model. *J. Neural Eng.* **11**, 066005 (2014).
57. Schaeffer, M.-C. & Aksenova, T. Switching Markov decoders for asynchronous trajectory reconstruction from ECoG signals in monkeys for BCI applications. *J. Physiol.-Paris* **110**, 348–360 (2016).
58. Maleki, M., Manshouri, N. & Kayikçioğlu, T. Fast and accurate classifier-based brain-computer interface system using single channel EEG data. in *2018 26th Signal Processing and Communications Applications Conference (SIU)* 1–4 (2018). doi:10.1109/SIU.2018.8404376.
59. Trejo, L. J., Rosipal, R. & Matthews, B. Brain-computer interfaces for 1-D and 2-D cursor control: designs using volitional control of the EEG spectrum or steady-state visual evoked potentials. *IEEE Trans. Neural Syst. Rehabil. Eng.* **14**, 225–229 (2006).
60. Wold, S., Ruhe, A., Wold, H. & Dunn, W. J., III. The Collinearity Problem in Linear Regression. The Partial Least Squares (PLS) Approach to Generalized Inverses. *SIAM J Sci Stat Comput* **5**, 735–743 (1984).
61. Eliseyev, A. & Aksenova, T. Recursive N-way partial least squares for brain-computer interface. *PloS One* **8**, e69962 (2013).
62. Dayal, B. S. & MacGregor, J. F. Recursive exponentially weighted PLS and its applications to adaptive control and prediction. *J. Process Control* **7**, 169–179 (1997).

63. Dayal, B. S. & MacGregor, J. F. Improved PLS algorithms. *J. Chemom.* **11**, 73–85 (1997).
64. Bro, R. Multi-way analysis in the food industry: models, algorithms, and applications. (Københavns UniversitetKøbenhavns Universitet, LUKKET: 2012 Det Biovidenskabelige Fakultet for Fødevarer, Veterin\a ermedicin og NaturressourcerFaculty of Life Sciences, LUKKET: 2012 Institut for FødevarevidenskabDepartment of Food Science, LUKKET: 2012 Kvalitet og TeknologiQuality & Technology, 1998).
65. Bro, R. Multiway calibration. Multilinear PLS. *J. Chemom.* **10**, 47–61 (1996).
66. Cichocki, A. *et al.* Tensor Decompositions for Signal Processing Applications: From two-way to multiway component analysis. *IEEE Signal Process. Mag.* **32**, 145–163 (2015).
67. Bishop, C. M. *Pattern Recognition and Machine Learning*. (Springer New York, 2006).
68. Schaeffer, M.-C. ECoG signal processing for Brain Computer Interface with multiple degrees of freedom for clinical application. (Université Grenoble Alpes, 2017).
69. Sutton, C. & McCallum, A. An Introduction to Conditional Random Fields. *Found. Trends Mach. Learn.* **4**, 267–373 (2012).
70. Rabiner, L. R. A tutorial on hidden Markov models and selected applications in speech recognition. *Proc. IEEE* **77**, 257–286 (1989).
71. Brain Computer Interface: Neuroprosthetic Control of a Motorized Exoskeleton ClinicalTrials.gov. https://clinicaltrials.gov/ct2/show/NCT02550522.
72. ICTRP clinical trial NCT02550522. https://apps.who.int/trialsearch/Trial2.aspx?TrialID=NCT02550522.
73. Morinière, B., Verney, A., Abroug, N., Garrec, P. & Perrot, Y. EMY: a dual arm exoskeleton dedicated to the evaluation of Brain Machine Interface in clinical trials. in *2015 IEEE/RSJ International Conference on Intelligent Robots and Systems (IROS)* 5333–5338 (2015). doi:10.1109/IROS.2015.7354130.
74. Volkova, K., Lebedev, M. A., Kaplan, A. & Ossadtchi, A. Decoding Movement From Electrocorticographic Activity: A Review. *Front. Neuroinformatics* **13**, (2019).
75. Nakanishi, Y. *et al.* Mapping ECoG channel contributions to trajectory and muscle activity prediction in human sensorimotor cortex. *Sci. Rep.* **7**, 1–13 (2017).
76. Young, D. *et al.* Closed-loop cortical control of virtual reach and posture using Cartesian and joint velocity commands. *J. Neural Eng.* **16**, 026011 (2019).
77. Farrokhi, B. & Erfanian, A. A piecewise probabilistic regression model to decode hand movement trajectories from epidural and subdural ECoG signals. *J. Neural Eng.* **15**, 036020 (2018).
78. Fukuma, R. *et al.* Closed-Loop Control of a Neuroprosthetic Hand by Magnetoencephalographic Signals. *PLOS ONE* **10**, e0131547 (2015).
79. Jerbi, K. *et al.* Inferring hand movement kinematics from MEG, EEG and intracranial EEG: From brain-machine interfaces to motor rehabilitation. *IRBM* **32**, 8–18 (2011).
80. Nguyen, C. H., Karavas, G. K. & Artemiadis, P. Adaptive multi-degree of freedom Brain Computer Interface using online feedback: Towards novel methods and metrics of mutual adaptation between humans and machines for BCI. *PLOS ONE* **14**, e0212620 (2019).
81. Vidaurre, C., Schlogl, A., Cabeza, R., Scherer, R. & Pfurtscheller, G. A fully on-line adaptive BCI. *IEEE Trans. Biomed. Eng.* **53**, 1214–1219 (2006).

**Acknowledgements**

CLINATEC is a Laboratory of CEA-Grenoble and has statutory links with the University Hospital of Grenoble (CHUGA) and with University Grenoble Alpes (UGA). This study was funded by CEA (recurrent funding) and the French Ministry of Health (Grant PHRC-15-15-0124), Institut Carnot, Fonds de Dotation CLINATEC. Fondation Philanthropique Edmond J Safra is a major founding institution of the CLINATEC Edmond J Safra Biomedical Research Center.

We thank Marion Mainsant, Antoine Lassauce, Gaël Reganha, Vincent Rouanne, Benoit Milville, Jean-Claude Royer, Stéphane Pezzani and the CHUGA members for the help in the experimental setup and clinical trial experiments.

**Author Contributions**

A.M. and T.A. developed and implemented the algorithm and software and wrote the manuscript. T.C. did the experiments. S.K., A.M. and F.M. implemented the algorithm to the exoskeleton platform, C.L. analyzed the ECoG data at rest, G.C. supervised the design and utilization of the recording platform, A.V. designed the exoskeleton, S.C. and A.L.B. did the surgery, A.L.B. supervised the Brain Computer Interface CLINATEC program, all authors reviewed the manuscript.

**Additional Information**

**Declaration of interests**

We declare no competing interests.

**Supplementary materials**

Supplementary videos SV1, SV2 and SV3 present examples of sessions 36,106,167 days after the last model calibration using exoskeleton effector.